\documentclass[10pt,draft,onecolumn]{IEEEtran}
\usepackage{amsmath,amssymb,threeparttable,placeins,enumerate,caption}
\usepackage{mathtools,relsize,cite}
\usepackage[pdftex]{graphicx}
\usepackage{epstopdf}
\usepackage[usenames]{color}
\usepackage{amsfonts}
\usepackage{latexsym}
\usepackage{subfigure,multirow,makecell,stfloats}

\definecolor{burntorange}{rgb}{0.8, 0.33, 0.0}

\usepackage{algorithm}
\usepackage[noend]{algpseudocode}

\newtheorem{theorem}{{{\textit{Theorem}}}}

\newtheorem{lemma}{{{\textit{Lemma}}}}

\newtheorem{property}{{{\textit{Property}}}}
\newtheorem{definition}{{{\textit{Definition}}}}

\newtheorem{remark}{{{\textit{Remark}}}}

\newtheorem{example}{{{\textit{Example}}}}
\newtheorem{result}{{{\textit{Result}}}}

\begin{document}
	
	
	\title{Constructions of Cross Z-Complementary Pairs with New Lengths}
	\author{ Avik Ranjan Adhikary, Zhengchun Zhou,
	Yang Yang, and Pingzhi Fan.
	\thanks{
%

Avik Ranjan Adhikary and Zhengchun Zhou  are with the School of Mathematics, Southwest Jiaotong University,
Chengdu, 611756, China, and also with the State
Key Laboratory of Cryptology, Beijing 100878, China. E-mail: Avik.Adhikary@ieee.org, zzc@swjtu.edu.cn}
\thanks{Yang Yang is with the School of Mathematics, Southwest Jiaotong University,
Chengdu, 611756, China. E-mail:  yang\_data@swjtu.edu.cn}
\thanks{Pingzhi Fan is with the Provincial Key Lab of Information Coding and Transmission, Southwest Jiaotong University,
Chengdu, 611756, China. E-mail: p.fan@ieee.org}
}
\maketitle
	
	\begin{abstract}
		Spatial modulation (SM) is a new paradigm of multiple-input multiple-output (MIMO) systems, in which only one antenna at the transmitter is activated during each symbol period. Recently, it is observed that SM training sequences derived from corss Z-complementary pairs (CZCPs) lead to
		optimal channel estimation performance over frequency-selective
		channels. CZCPs are special form of sequence pairs which have zero aperiodic autocorrelation zones and cross-correlation zone at certain time-shifts. Recent paper by Liu \textit{et al.} discussed only perfect CZCPs. In this paper, we focus on non-perfect CZCPs. We introduce the term cross Z-complementary ratio and re-categorise the CZCPs, both perfect and non-perfect, based on that. We propose a systematic construction of CZCPs based on generalized Boolean functions (GBFs). We further extend the lengths of the CZCPs by using the insertion method. The proposed CZCPs are all of new lengths of the form $2^\alpha10^\beta26^\gamma+2~(\alpha\geq1)$, $10^\beta+2$, $26^\gamma+2$ and $10^\beta 26^\gamma+2$. Finally we propose a construction of optimal binary CZCPs having parameters $(12,5)$ and $(24,11)$ from binary Barker sequences. These CZCPs are also extended to $(12N,5N)$- CZCPs and $(24N,11N)$- CZCPs, where $N$ is the length of a binary Golay complementary pair (GCP). During the proof, we also found a new structural property of binary CZCPs and concluded all binary GCPs are also CZCPs. Finally, we give some numerical simulations to confirm that depending on the number of multi-paths, the proposed CZCPs can be used to design SM training matrix which attains the minimum mean square error.
	\end{abstract}

\vspace{-0.2cm}
	
	\section{Introduction}
	\IEEEPARstart{I}{n} 1950, M. J. Golay introduced complementary pairs in his work on multislit spectrometry \cite{golay1}. Golay complementary pairs (GCPs) are pair of sequences whose aperiodic autocorrelation sums (AACSs) are zero everywhere, except at the zero shift \cite{golay2}. Binary GCPs are available only for limited lengths of the form $2^\alpha10^\beta26^\gamma$ (where $\alpha$, $\beta$, and $\gamma$ are non-negative integers)\cite{golay2,Borwein03}. In 1972, Tseng and Liu \cite{tseng72} extended the idea of complementary pairs to complementary sets (CSs) of sequences. Since then, CSs found a number of applications in communication systems \cite{spasojevic,davis99,paterson,popovic,schmidt}. Due to the limited availability of binary GCPs Fan \textit{et al.} \cite{fan} proposed binary Z-complementary pairs (ZCPs) in 2007. Since then, a lot of research has been done afterwords towards the systematic and structural analysis of ZCPs \cite{li,zilongobzcp,zilongebzcp,chen17,Avik_iwsda,Avik_tit,Avik_ebzcp}.
	
	
	Spatial modulation (SM) is a special kind of multiple-input multiple-output (MIMO) technique, which optimizes multiplexing gain with complexity and performance \cite{Mesleh1,Mesleh2,renzo,yang,wen}. The main difference of SM system with a traditional MIMO is that it is equipped with a single radio-frequency (RF) chain. In SM, only one transmit antenna is activated over every symbol duration. During each time-slot, an SM symbol can be divided into two parts, spatial symbol and constellation symbol. Spatial symbol is responsible for the transmit antenna elements and constellation symbol is selected from a conventional phase shift keying (PSK)/quadrature amplitude modulation (QAM) constellation and transmitted from the active transmit antenna element. Such unique transmission
	principle of SM allows it to have the salient advantages of
	zero inter-channel interference, low energy consumption \cite{yang1}, and
	low receiver complexity over traditional MIMO systems. Till date, however, little has been understood on channel estimation of SM in frequency-selective channels. Early literature on SM mostly assume that channel state information (CSI) is perfectly known at the receiver \cite{Jeganathan,Renzo12}. Note that the ``one-RF-chain" principle of SM prevents the
	transmitter from using simultaneous pilot transmission over all
	the transmit antennas. Consequently, it implies that dense training sequences proposed in \cite{yang3,Fragouli,fan2} for traditional MIMO are not applicable in SM systems. Although an identity training matrix has been employed for joint channel estimation and data detection in SM systems \cite{Sugiura}, extension to frequency-selective channels
	is not straightforward. A naive scheme is to extend a perfect
	sequence (having zero autocorrelation sidelobes) with cyclic prefix
	(CP) and then send the extended sequence in turn over
	multiple transmit antennas. But this training scheme would be inefficient in highly dispersive channels.
	
	To deal with this problem, recently Liu \textit{et al.} \cite{zilong_ccp} proposed a new class of sequence pairs called cross Z-complementary sequence pairs (CZCPs). The authors also proposed a generic training framework for SM training
	over frequency selective channels. Under the proposed framework in \cite{zilong_ccp}, the authors derived the lower bound on channel estimation mean square error
	(MSE) using least square (LS) estimator and conditions to
	meet the lower bound with equality. In \cite{zilong_ccp}, the authors show that CZCPs play an instrumental role in the design of optimal SM training sequences (which are equivalent to certain
	sparse matrices). The authors also show that the numerical simulations indicate that
	the proposed SM training sequences in \cite{zilong_ccp} lead to minimum
	channel estimation MSE w.r.t. the aforementioned lower
	bound.

		\subsection{Concept of Cross Z-Complementarity}
	Let $\mathbf{a}$ and $\mathbf{b}$ be two sequences of length $N$. Also let $\rho_{\mathbf{a},\mathbf{b}}(\tau)$ denotes the aperiodic cross-correlation function of $\mathbf{a}$ and $\mathbf{b}$ at time-shift $\tau$ given by
		\begin{equation}
	\rho_{\mathbf{a},\mathbf{b}}(\tau):= \left \{
	\begin{array}{cl}
	\sum\limits_{k=0}^{N-1-\tau}\omega^{a_k-b_{k+\tau}},&~~0\leq \tau \leq N-1;\\
	\sum\limits_{k=0}^{N-1+\tau}\omega^{a_{k-\tau}-b_k},&~~-(N-1)\leq \tau \leq -1;\\
	0,& ~~\mid \tau \mid \geq N,
	\end{array}
	\right .
	\end{equation}
	where $\omega=\exp(2\pi\sqrt{-1}/q)$ ($q \geq 2$, is a positive integer). When the two sequences are identical, i.e., $\mathbf{a} = \mathbf{b}$, $\rho_{\mathbf{a},\mathbf{b}}(\tau)$ is known as an aperiodic auto-correlation function (AACF) of $\mathbf{a}$ and it is denoted by $\rho_{\mathbf{a}}(\tau)$. Then $(\mathbf{a},\mathbf{b})$ is said to be a $(N,Z+1)$- Z-complementary pair (ZCP) if
	\begin{equation}
	\rho_{\mathbf{a}}(\tau)+\rho_{\mathbf{b}}(\tau)=0, \text{ for }1\leq \tau \leq Z.
	\end{equation}
	ZCPs are related to the AACS only. To define CZCPs we need the following two sets. For an integer $Z$, let $\mathcal{T}_1=\{1,2,\dots,Z\}$ and $\mathcal{T}_2=\{N-Z,N-Z+1,\dots,N-1\}$. Then $(\mathbf{a},\mathbf{b})$ is called an $(N,Z)$- CZCP if it possesses symmetric zero (out-of-phase) AACSs for time shifts over $\mathcal{T}_1 \cup \mathcal{T}_2$ and zero aperiodic cross-correlation sums (ACCSs) for time shifts over $\mathcal{T}_2$ \cite{zilong_ccp}. In short, it needs to satisfy the following two conditions:
	\begin{equation}\label{con_eq}
	\begin{split}
	&\text{C1: }\rho_{\mathbf{a}}(\tau)+\rho_{\mathbf{b}}(\tau)=0, \text{for all }|\tau| \in \mathcal{T}_1 \cup \mathcal{T}_2; \text{ and}\\
	&\text{C2: }\rho_{\mathbf{a},\mathbf{b}}(\tau)+\rho_{\mathbf{b},\mathbf{a}}(\tau)=0, \text{for all }|\tau| \in \mathcal{T}_2.
	\end{split}
	\end{equation}
	From C1 it is clear that CZCPs have two zero auto-correlation zones (ZACZs) when the AACSs are considered. In this paper we will call them ``front-end-ZACZ" and ``tail-end-ZACZ" for the time shifts over $\mathcal{T}_1$ and $\mathcal{T}_2$, respectively. From C2, we get that each CZCP needs to have ``tail-end zero cross-correlation zone (ZCCZ)" when ACCS are considered.
	
	In \cite{zilong_ccp}, Liu \textit{et al.} proposed the concept of cross Z-complementarity and constructed a class of GCPs of length $N$ achieving the maximum ZACZ and ZCCZ width of $N/2$. The authors in \cite{zilong_ccp} termed them as ``strengthened GCPs" and the $(N,Z)$- CZCPs for which the value of $Z$ is $N/2$, as perfect CZCPs. On contrary, the $(N,Z)$- CZCPs for which $Z<N/2$, the authors termed them as non-perfect CZCPs. Since, $Z=N/2$ can only be achievable when $N$ is the length of a complementary pair, we re-categorise the CZCPs based on cross Z-complementary ratio.

	\subsection{Cross Z-Complementary Ratio}
\begin{figure}
		\centering
		\includegraphics[width=\columnwidth,draft=false]{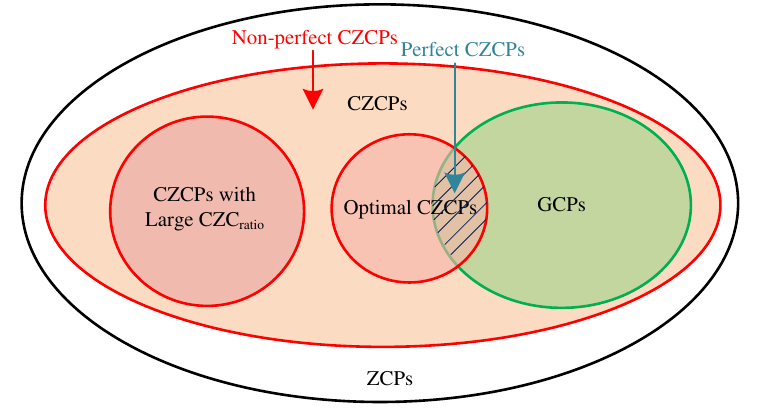}
		\caption{Relationship of binary CZCPs with binary ZCPs and binary GCPs. \label{f2}}
	\end{figure}

	Let $(\mathbf{a},\mathbf{b})$ be a CZCP of length $N$ having ZCCZ and ZACZ of length $Z$. If the maximum achievable length of $Z$ is $Z_{\max}$, then we define the Cross Z-Complementary ratio ($CZC_{ratio}$) as
	\begin{equation}
		CZC_{ratio}=\frac{Z}{Z_{\max}}.
	\end{equation}
	When $CZC_{ratio}=1$, we call it optimal. Systematic construction of optimal CZCPs is very challenging, specially for binary cases, when the length $N$ is not of the form $2^\alpha10^\beta26^\gamma$ and remains open. Although
	two binary optimal CZCPs of lengths $12$ and $24$ have been found in this paper, they are constructed via Barker sequences which are available only for lengths up to 13. Alternatively, we aim for a systematic construction for CZCPs with large $CZC_{ratio}$, e.g., when $0.5<CZC_{ratio}$. In practice, a large $CZC_{ratio}$ means that more multi-path can be supported in the SM enabled MIMO transmission \cite{zilong_ccp}. Figure \ref{f2} clearly categorizes the CZCPs and also shows the relation of CZCPs with the existing GCPs and ZCPs. Figure \ref{f2} also corrects \cite[Figure 3]{zilong_ccp}, since there are a lot of ``non-strengthened GCPs", where $Z<N/2$, which were not shown in \cite[Figure 3]{zilong_ccp}. The cross-correlation property between the sequences of binary GCP satisfies the conditions to be a binary CZCP, with $Z=1$. We have shown it in Theorem \ref{th_new7}.
	
	\subsection{Our Contributions}

Based on the simulations given in \cite{zilong_ccp}, it can be observed that CZCPs can be used as an alternative to strengthened GCPs in the design of optimal SM training sequences over frequency selective channels. Since the availability of binary GCPs only for lengths of the form $2^\alpha 10^\beta 26^\gamma$ is still a conjecture, from \cite[Table I]{zilong_ccp} it can be partially proved that the maximum ZACZ and ZCCZ for binary non-perfect CZCPs of length $N$ is $N/2-1$. In \cite{zilong_ccp}, the authors proposed two constructions of perfect $(N,N/2)$- CZCPs, where for binary case $N=2^{\alpha+1} 10^\beta 26^\gamma$, $\alpha,\beta,\gamma\in\mathbb{Z}^+,\alpha\geq1$. Motivated by the work of Liu \textit{et al.} \cite{zilong_ccp}, we propose several new constructions of $(N,Z)$- CZCPs of new lengths. Specifically, the contribution of this paper are the following:
\begin{itemize}
	\item We propose construction of ($2^{m-1}+2,2^{\pi(m-3)}+1$)- CZCPs by using generalized Boolean functions (GBFs), where $m\geq 4$ and $\pi$ is a permutation over $\{0,1,\dots,m-3\}$.
	\item We further extend the construction by applying insertion method on GCPs, which are constructed via Turyn's method. By exploiting the intrinsic structural properties of the GCPs found in \cite{Avik_tit}, we propose systematic construction of CZCPs of new lengths of the form $2^\alpha 10^\beta 26^\gamma~(\alpha\geq1)+2$, $10^\beta+2$, $26^\gamma+2$ and $10^\beta26^\gamma+2$ based on insertion method. The constructions which are based on insertion method, all the GCPs are constructed by applying Turyn's method over kernel GCPs given in Table \ref{table_new}.
	\renewcommand{\arraystretch}{0.6}
	\begin{table}
		\small
		\centering
		\tabcolsep=0.11cm
		\caption{\cite{Borwein03} GCP Kernels of Lengths $2$, $10$ and $26$. \label{table_new}}
		\resizebox{\columnwidth}{!}{
			
			\begin{tabular}{|c||c||c|}
				\hline
				$N$ & $\left( \begin{matrix}
				\mathbf{a}\\
				\mathbf{b}
				\end{matrix} \right)$ & Notation   \\ \hline \hline
				$2$  & $\left( \begin{matrix}
				++\\
				+-
				\end{matrix} \right) $ & $K_2$  \\  \hline
				$10$ & $ \left ( \begin{matrix}
				++-+-+--++\\
				++-+++++--
				\end{matrix} \right) $ & $K_{10}$  \\  \hline
				$26$ & $\left ( \begin{matrix}
				++++-++--+-+-+--+-+++--+++\\
				++++-++--+-+++++-+---++---
				\end{matrix} \right) $ & $K_{26}$  \\  \hline
			\end{tabular}
		}
	\end{table}
	\item We propose an optimal construction of binary $(12,5)$- CZCP and $(24,11)$- CZCP using binary Barker sequences. These two optimal CZCPs leads to two new sets of CZCPs with parameters $(12N,5N)$ and $(24N,11N)$, where $N$ is the length of a GCP.
	\item Along with this, we also found that all the binary GCPs are also CZCPs. We clearly describe the relationships between binary CZCPs with ZCPs and GCPs in Figure \ref{f2}.
	\item Through numerical simulations we show that our proposed optimal and almost-optimal CZCPs can be used to design optimal training sequences for SM systems, based on the framework given in \cite{zilong_ccp}.
	\item We analysed through numerical simulations that although some of the CZCPs may not be optimal or almost-optimal in sequence design point of view, however, depending on the number of multi-paths, those CZCPs can still be used to design optimal training sequences for SM systems.
\end{itemize}

 Based on the discussion on designing training sequences for broadband SM systems and the numerical simulations afterwards in \cite{zilong_ccp}, it can be realised that our proposed constructions will add flexibility in choosing the CZCPs of various lengths for designing training sequences in SM systems. Since, in practical scenarios, a longer training sequence will give rise to a higher
 training overhead, therefore, selection of the training length is a trade-off between channel estimation performance and training overhead. For example, let us consider that in a practical scenario, $22$ multi-path is required. If only the CZCPs in \cite{zilong_ccp} are considered, then one have to use a $(64,32)$- CZCP. However, by our proposed constructions, $(48,22)$- CZCP can be used to design the training sequences in SM systems using the framework given in \cite{zilong_ccp}. This will improve the system performance.

The rest of the paper is organised as follows. In Section II, some previous results on binary GCPs and also recall the basics of GBFs is discussed. In Section III, ($2^{m-1}+2,2^{\pi(m-3)}+1$)- CZCPs by using GBFs are proposed. In Section IV, the construction of CZCPs of various lengths of the form $2^\alpha 10^\beta 26^\gamma+2~(\alpha \geq1)$, $10^\beta+2$, $26^\gamma+2$ and $10^\beta 26^\gamma+2$ via insertion method are proposed. Optimal binary $(12,5)$- CZCP and $(24,11)$- CZCP from Barker sequences are proposed in Section IV. $(12N,5N)$- CZCPs and $(24N,11N)$- CZCPs are constructed in Section V. In Section VI, the performance of the proposed CZCPs to design sparse training matrices for optimal channel estimation performance (w.r.t. the derived lower bound in \cite{zilong_ccp}) in SM-MIMO
frequency selective channels, are analysed. Finally, we conclude the paper in Section VII.

	\section{Preliminary Results}
	Let us fix the following notations, which will be used throughout the paper.
	\begin{itemize}
		\item $\overleftarrow{\mathbf{a}}$ denotes the reverse of the sequence $\mathbf{a}$.
		\item $1$ and $-1$ are denoted by $+$ and $-$, respectively.
		\item $\mathbf{c}_L$ denotes a length-$L$ vector with identical entries of $c$.
		\item $\forall$ denotes `for all'.
		\item $\bar{x}=1-x$ is the binary complement of $x\in \{0,1\}$.
		\item $\lfloor.\rfloor$ denotes the floor function.
		\item $\mathbb{U}_q=\{exp\frac{2\pi\sqrt{-1}t}{q}:0\leq t <q\}$ denotes the set of $q$-th roots of complex numbers.
		\item $X^T$ and $X^H$ denote the transpose and the Hermitian
		transpose of matrix $X$, respectively.
		\item $Tr(X)$ denotes the trace of square matrix $X$.
		\item $\mathbf{r}_i\equiv(r_{i,0},r_{i,1},\dots, r_{i,m-1})$ is the binary vector representation of integer $i$ $(i=\displaystyle \sum_{j=0}^{m-1}r_{i,j}2^j)$.
		\item $\omega=exp\frac{2\pi\sqrt{-1}}{q}$ denotes the $q$-th root of unity.
	\end{itemize}

	\begin{definition}[Complementary Mates]
		A GCP $(\mathbf{c},\mathbf{d})$ is called a complementary mate of GCP $(\mathbf{a},\mathbf{b})$ if
		\begin{equation}
		\rho_{\mathbf{a},\mathbf{c}}(\tau)+\rho_{\mathbf{b},\mathbf{d}}(\tau)=0, \text{ for all }0\leq \tau<N.
		\end{equation}
		Throughout the paper, for a GCP $(\mathbf{a},\mathbf{b})$ we consider $(\mathbf{c},\mathbf{d})\equiv (\overleftarrow{\mathbf{b^*}},-\overleftarrow{\mathbf{a^*}})$ as its complementary mate.
\end{definition}

\begin{definition}[Optimal and Almost-optimal CZCPs]
	We define a $(N,Z)$- CZCP $(\mathbf{a},\mathbf{b})$ to be optimal if $CZC_{ratio}=1$ and almost optimal if the $CZC_{ratio}$ is near to $1$.
	
	
	Note that all perfect CZCPs mentioned in \cite{zilong_ccp} are optimal. However, non-perfect CZCPs can also be optimal. Fig. \ref{f2} explains the definitions clearly.
\end{definition}
	
	
	\begin{definition}[Symmetric Insertion Function] Consider a sequence $\mathbf{a}$ of even length $N$, given by $(a_0,a_1, \dots ,a_{N-1})$. Then $\mathcal{I}_s(\mathbf{a},r,\{x_0,x_1\})$, given by
		\begin{equation}\label{definsertion}
		\begin{split}
		&\mathcal{I}_s(\mathbf{a},r,\{x_0,x_1\}) =\\
		&\begin{cases}
		(x_0,a_0,a_1, \dots ,a_{N-1},x_1), & \text{if }r=0, \\
		(x_1,a_0,a_1, \dots ,a_{N-1},x_0), & \text{if }r=N, \\
		(a_0,a_1, \dots,a_{N/2-1},x_0,x_1, a_{N/2},\\\hspace{3cm}\dots ,a_{N-1}), & \text{if }r=N/2,\\
		(a_0,a_1, \dots ,a_{r-1},x_0,a_r, \\\hspace{0.5cm}\dots,a_{N-r-1},x_1,a_{N-r},\dots ,a_{N-1}), & \text{if }0\leq r\leq N,\\&\hspace{0.1cm}~r\neq N/2 \\
		\end{cases}
		\end{split}
		\end{equation}
		is an insertion function which generates sequence of length $N+2$ with element $x_0$ at the $r$-th position and element $x_1$ at the $(N-r)$-th position when $r\neq N/2$. For $r=N/2$, $x_0$ and $x_1$ is inserted at $N/2$-th and $(N/2+1)$-th position, respectively.
	\end{definition}

\subsection{Generalized Boolean Functions (GBFs)}

A GBF $f:\mathbb{Z}_2^m\rightarrow\mathbb{Z}_{q}$ can uniquely be written as a linear combination of $2^m$ monomials
\begin{equation}
\begin{split}
1,x_0,x_1,\cdots,x_{m-1},&x_0x_1,x_0x_2,\cdots,\\&x_{m-2}x_{m-1}\cdots, x_0x_1\cdots x_{m-1},
\end{split}
\end{equation}
where the coefficients are taken from $\mathbb{Z}_{q}$.

By the notation $\Psi(f)$, we denote the sequence corresponding to a GBF $f$ and defined by $(\omega^{f_0}, \omega^{f_1}, \cdots, \omega^{f_{2^m-1}})$ where $f_i=f(r_{i,0},r_{i,1},\dots, r_{i,m-1})$ and $\mathbf{r}_i\equiv(r_{i,0},r_{i,1},\dots, r_{i,m-1})$ is the binary vector representation of integer $i$ $(i=\displaystyle \sum_{j=0}^{m-1}r_{i,j}2^j)$.


Given a GBF $f$ with $m$ variables, as defined above, the corresponding sequence $\Psi(f)$ will be of length $2^m$. $\mathbf{1}$ denotes a sequence of all ones. In this paper, we concern about $(N,Z)$- CZCPs, where $N \neq 2^m$. Hence we define the truncated sequence $\Psi_{L}(f)$ corresponding to GBF $f$ by eliminating the first and last $L$ elements of the sequence $\Psi(f)$.

\begin{example}
	Let us consider $m=3$, $q=2$ and $f=x_0x_1+x_1x_2$, then
	\begin{equation}
	\begin{split}
	x_0x_1&=(0,0,0,1,0,0,0,1),\\
	x_1x_2&=(0,0,0,0,0,0,1,1),\\
	x_0x_1+x_1x_2&=(0,0,0,1,0,0,1,0),
	\end{split}
	\end{equation}
	and therefore $\Psi(f)=(+++-++-+)$. Let $L=1$, then $\Psi_{1}(f)=(++-++-)$.
\end{example}

\begin{remark}
	Note that in \cite{chen2016} only the last $L$-bits were truncated. In our work we are eliminating the first as well as the last $L$-bits of the sequence.
\end{remark}

\begin{lemma}[\cite{rathinakumar}]\label{lem3}
	Let $\pi$ be a permutation of $\{0,1,2,\dots, m-1\}$. Consider a GBF $f:\mathbb{Z}_2^m\rightarrow \mathbb{Z}_{2}$, given by
	\begin{equation}
	f=\sum_{\alpha=0}^{m-2}x_{\pi(\alpha)}x_{\pi(\alpha+1)}+\sum_{i=0}^{m-1}g_ix_i+g^\prime,
	\end{equation}
	and
	\begin{equation}
	\bar{f}=\sum_{\alpha=0}^{m-2}\bar{x}_{\pi(\alpha)}\bar{x}_{\pi(\alpha+1)}+\sum_{i=0}^{m-1}g_i\bar{x}_i+g^\prime,
	\end{equation}
\end{lemma}
where $g_i,g^\prime \in \mathbb{Z}_2$. Then, $(\Psi(\bar{f}+\bar{x}_{\pi(m-1)}),\Psi(\bar{f}+\mathbf{1}))$ is one of the complementary mates of $(\Psi(f),\Psi(f+x_{\pi(m-1)}))$.
%
	\begin{lemma}[Turyn's Method \cite{Turyn74}]\label{thm_turyn}
	Let $\mathcal{A}=(\mathbf{a},\mathbf{b})$ and $\mathcal{B}= (\mathbf{c},\mathbf{d})$ be binary GCPs of lengths $N$ and $M$, respectively and denote $\mathcal{A}$ as the $1st$ pair and $\mathcal{B}$ as the $2nd$ pair. Then $(\mathbf{e},\mathbf{f})\triangleq Turyn(\mathcal{A},\mathcal{B})$ is a GCP of length-$MN$, where
	\begin{equation} \label{turyn_1st_form}
	\begin{split}
	\mathbf{e} & =\mathbf{c} \otimes \left(\mathbf{a}+\mathbf{b} \right) / 2 - \overleftarrow{\mathbf{d}} \otimes \left(\mathbf{b}-\mathbf{a} \right) / 2,  \\ \mathbf{f} & =\mathbf{d} \otimes \left( \mathbf{a}+\mathbf{b} \right) / 2 + \overleftarrow{\mathbf{c}} \otimes  \left( \mathbf{b}-\mathbf{a} \right) / 2.
	\end{split}
	\end{equation}
\end{lemma}

	\begin{result}\cite{Avik_tit}\label{r1}
	Let $\mathcal{A} = (\mathbf{a},\mathbf{b})$ be a binary GCP kernel $K_N $ where $N\in \{2,10,26\}, \mathcal{B} = (\mathbf{c},\mathbf{d})$ be a GCP of length $M$ and $(\mathbf{e},\mathbf{f})=Turyn(\mathcal{A},\mathcal{B})$. If the $i$-th column of $\mathcal{B}$ has elements with the same sign, then the next $N$ columns of $(\mathbf{e},\mathbf{f})$ will have elements with same sign, starting from $Ni$-th column. If the $i$-th column of $\mathcal{B}$ have elements with different signs, then the next $N$ columns of $(\mathbf{e},\mathbf{f})$ will have elements with different sign, starting from $Ni$-th column.
\end{result}

	\begin{result}\cite{Avik_tit}\label{r2}
	Let $(\mathbf{e},\mathbf{f})$ be a GCP of length $2^\alpha M$, constructed recursively using Turyn's method on kernel GCPs as follows:
	\begin{equation}\label{fac_2_tur_equ}
	\begin{split}
	& (\mathbf{e}_0,\mathbf{f}_0)=K_2, \\
	& (\mathbf{e}_i,\mathbf{f}_i)=Turyn(\mathcal{A},(\mathbf{e}_{i-1},\mathbf{f}_{i-1})), ~ \mathcal{A}=K_2,~K_{10} \text{ or } K_{26},
	\end{split}
	\end{equation}
	where $M=10^\beta 26^\gamma$ and $\alpha,~\beta, \text{ and } \gamma$ are non-negative integers and $\alpha \geq 1$. Then the first $2^{\alpha -1} M$ columns of $(\mathbf{e},\mathbf{f})$ will have elements with identical sign in each column.
\end{result}

\begin{result}\cite{Avik_tit}\label{r3}
	Let $(\mathbf{e},\mathbf{f})$ be a GCP of length $10^\beta$ or $26^\gamma$, constructed iteratively using Turyn's method on $K_{10}$ or $K_{26}$, respectively. Also suppose there are $t$ consecutive columns of $K_{10}$ or $K_{26}$, having elements with identical signs (or different signs) in each column, starting from the $i$-th column index. Then, the $t\times 10^{\beta-1}$ or $t\times 26^{\gamma-1}$ consecutive columns of $(\mathbf{e},\mathbf{f})$ will have elements with identical sign (or different sign) in each column, starting from the $iN^{p-1}$-th column, respectively.
\end{result}

\begin{result}\cite{Avik_tit}\label{r4}
	Let $(\mathbf{e},\mathbf{f})$ be a GCP of length $10^\beta 26^\gamma$, constructed iteratively by employing Turyn's method on $K_{10} \text{ and }K_{26}$ as follows:
	\begin{equation}\label{fac_2_tur_equ2}
	\begin{split}
	& (\mathbf{e}_0,\mathbf{f}_0)=K_{26}, \\
	& (\mathbf{e}_i,\mathbf{f}_i)=Turyn(\mathcal{A},(\mathbf{e}_{i-1},\mathbf{f}_{i-1})), ~ \mathcal{A}=K_{10} \text{ or } K_{26},
	\end{split}
	\end{equation}
	where $\beta \text{ and } \gamma$ are non-negative integers. Then the first $12\times 26^{\gamma -1} 10^\beta$ columns of $(\mathbf{e},\mathbf{f})$ will have elements with identical signs in each column.
\end{result}

In the following section we propose another construction of CZCPs with the help of GBFs.

	\section{Construction of CZCPs through GBFs}
The proposed construction is discussed in this subsection. We need the following lemmas for the construction.
\begin{lemma}\label{new_lem}
	Let $r_{{2^{m-2}+\tau-1},\pi(m-3)}$ and $r_{{3\times 2^{m-2}-\tau},\pi(m-3)}$ be as defined above. Then,
	\begin{equation}
	r_{{2^{m-2}+\tau-1},\pi(m-3)}+r_{{3\times 2^{m-2}-\tau},\pi(m-3)}=1,~ 0< \tau\leq 2^{m-2}.
	\end{equation}
\end{lemma}
\begin{IEEEproof}
	Note that binary representation of $x$ and $2^m-x-1$ for $0\leq x <2^m$ are always complementary to each other.
	Let $x=2^{m-2}+\tau-1$, then $2^m-(2^{m-2}+\tau-1)-1=3\times 2^{m-2}-\tau$. Hence, the binary representation of $2^{m-2}+\tau-1$ is complementary with $3\times 2^{m-2}-\tau$. Hence the proof follows.
\end{IEEEproof}

\begin{lemma}\label{lem4}
	For any integer $m\geq 4$, let $\pi$ be a permutation of $\{0,1,2,\dots, m-3\}$. Then, for $2^{m-1}-2^{\pi(m-3)}<\tau<2^{m-1}+1$, $r_{2^{m-2}+\tau-1,\pi(m-3)}$ is always $1$.
\end{lemma}
\begin{IEEEproof}
	Let $\pi(m-3)=v$. Also, let the binary representation of $i\leq 3\times 2^{m-2}-1$ be $(i_0,i_1,\dots,i_{m-3},0,1)$. If, additionally, $i\geq 3\times 2^{m-2} - 2^v$, then $i_s=1$ for $s=v,v+1,\ldots, m-3$. Therefore for $3\times 2^{m-2} - 2^{v}\leq i \leq 3\times 2^{m-2}-1$, $r_{i,v}=1$. Hence the proof follows.
	
\end{IEEEproof}

\begin{theorem}\label{th1_gbf}
	For any integer $m\geq 4$, let $\pi$ be a permutation of $\{0,1,2,\dots, m-3\}$. For $d\in \mathbb{Z}_2$, let the GBF $g^d:\mathbb{Z}_2^m\rightarrow \mathbb{Z}_{q}$ be given as follows:

\begin{equation}\label{gbf1}
g^d=\frac{q}{2} [\bar{x}_{m-1}x_{m-2}\zeta^d+x_{m-1}\bar{x}_{m-2}\eta^d+\bar{d}x_{m-1}x_{m-2}]+c
\end{equation}
	where $\zeta^d:\mathbb{Z}_2^{m-2}\rightarrow \mathbb{Z}_2$ is
	\begin{equation}\label{eq10_t}
	\zeta^d=\sum_{\alpha=0}^{m-4}x_{\pi(\alpha)}x_{\pi(\alpha+1)}+dx_{\pi(m-3)},
	\end{equation}
	$\eta^d:\mathbb{Z}_2^{m-2}\rightarrow \mathbb{Z}_2$ is
	\begin{equation}\label{eq11_t}
	\eta^d=\sum_{\alpha=0}^{m-4}\bar{x}_{\pi(\alpha)}\bar{x}_{\pi(\alpha+1)}+\bar{d}\bar{x}_{\pi(m-3)}+d\mathbf{1},
	\end{equation}
	and $c \in \mathbb{Z}_q$ for $0\leq i \leq m-3$.
	Then
	\begin{equation}
		(\mathbf{a},\mathbf{b})= \left(\Psi_{2^{m-2}-1}(g^0),\Psi_{2^{m-2}-1}(g^1)\right)
	\end{equation}
	forms a ($2^{m-1}+2,2^{\pi(m-3)}+1$)- CZCP.
\end{theorem}

\begin{IEEEproof}
 Using Lemma \ref{lem3}, since $(\Psi(\zeta^0),\Psi(\zeta^1))$ and $(\Psi(\eta^0),\Psi(\eta^1))$ are complementary mates of each other, therefore for $0< \tau<2^{m-2}$ we have
	\begin{equation}\label{eq12_t}
	\begin{split}
	&\rho_{\mathbf{a}}(\tau)+\rho_{\mathbf{b}}(\tau)\\
	&= \left[\omega^{-\frac{q}{2}\zeta^0(\mathbf{r}_{2^{m-2}+\tau-1})}+\omega^{-\frac{q}{2}+\frac{q}{2}\eta^0(\mathbf{r}_{3\times 2^{m-2}-\tau})}\right]\\
	& \hspace{0.2cm}+\left[\omega^{-\frac{q}{2}\zeta^1(\mathbf{r}_{2^{m-2}+\tau-1})}+\omega^{\frac{q}{2}\eta^1(\mathbf{r}_{3\times 2^{m-2}-\tau})}\right]\\
	&=\left[\omega^{-\frac{q}{2}\zeta^0(\mathbf{r}_{2^{m-2}+\tau-1})}+\omega^{-\frac{q}{2}\zeta^1(\mathbf{r}_{2^{m-2}+\tau-1})}\right]\\
	&\hspace{0.2cm}-\left[\omega^{\frac{q}{2}\eta^0(\mathbf{r}_{3\times 2^{m-2}-\tau})}-\omega^{\frac{q}{2}\eta^1(\mathbf{r}_{3\times 2^{m-2}-\tau})}\right].
	\end{split}
	\end{equation}
	Note that,
	\begin{equation}\label{eq13_t}
	\frac{q}{2}\zeta^1(\mathbf{r}_{2^{m-2}+\tau-1})=\frac{q}{2}\left[\zeta^0(\mathbf{r}_{2^{m-2}+\tau-1})+r_{2^{m-2}+\tau-1,\pi(m-3)}\right]
	\end{equation}
	and
	\begin{equation}\label{eq14_t}
	\frac{q}{2}\eta^0(\mathbf{r}_{3\times 2^{m-2}-\tau})=\frac{q}{2}\left[\eta^1(\mathbf{r}_{3\times 2^{m-2}-\tau})+r_{3\times 2^{m-2}-\tau,\pi(m-3)}\right].
	\end{equation}
	
	Applying (\ref{eq13_t}) and (\ref{eq14_t}) in (\ref{eq12_t}) we get
	\begin{equation}\label{eq15_t}
	\begin{split}
	&\rho_{\mathbf{a}}(\tau)+\rho_{\mathbf{b}}(\tau)\\
	&=\omega^{-\frac{q}{2}\zeta^0(\mathbf{r}_{2^{m-2}+\tau-1})}\left[1+\omega^{-\frac{q}{2}\times r_{2^{m-2}+\tau-1,\pi(m-3)}}\right]\\
	&\hspace{0.2cm}-\omega^{\frac{q}{2}\eta^1(\mathbf{r}_{3\times 2^{m-2}-\tau})}\left[\omega^{\frac{q}{2}\times r_{3\times 2^{m-2}-\tau,\pi(m-3)}}-1\right].
	\end{split}
	\end{equation}
	We have the following two sub-cases:
	\begin{itemize}
		\item[1)] Using Lemma \ref{new_lem}, for $r_{2^{m-2}+\tau-1,\pi(m-3)}=0$ we have $r_{3\times 2^{m-2}-\tau,\pi(m-3)}=1$. In this case it follows from (\ref{eq15_t}) that
		\begin{equation}
		\begin{split}
		&\rho_{\mathbf{a}}(\tau)+\rho_{\mathbf{b}}(\tau)\\
		&=2\left[\omega^{-\frac{q}{2}\zeta^0(\mathbf{r}_{2^{m-2}+\tau-1})}+\omega^{\frac{q}{2}\eta^1(\mathbf{r}_{3\times 2^{m-2}-\tau})}\right]\\
		&=0,
		\end{split}
		\end{equation}
		since, from (\ref{eq10_t}) and (\ref{eq11_t}) we get
		\begin{equation}
		\omega^{-\frac{q}{2}\zeta^0(\mathbf{r}_{2^{m-2}+\tau-1})}=-\omega^{\frac{q}{2}\eta^1(\mathbf{r}_{3\times 2^{m-2}-\tau})}.
		\end{equation}
		
		\item[2)] Using Lemma \ref{new_lem}, for $r_{2^{m-2}+\tau-1,\pi(m-3)}=1$ we have $r_{3\times 2^{m-2}-\tau,\pi(m-3)}=0$. In this case we have from (\ref{eq15_t})
		\begin{equation}
		\rho_{\mathbf{a}}(\tau)+\rho_{\mathbf{b}}(\tau)=0.
		\end{equation}
	\end{itemize}
	For $2^{m-1}-2^{\pi(m-3)}<\tau <2^{m-1}+1$ we have
	\begin{equation}\label{eq19_t}
	\begin{split}
	&\rho_{\mathbf{a}}(\tau)+\rho_{\mathbf{b}}(\tau)\\
	&=\left[\omega^{-\frac{q}{2}\eta^0(\mathbf{r}_{2^{m-2}+\tau-1})}+\omega^{-\frac{q}{2}+\frac{q}{2}\zeta^0(\mathbf{r}_{3\times 2^{m-2}-\tau})}\right]\\
	&\hspace{0.2cm}+\left[\omega^{-\frac{q}{2}\eta^1(\mathbf{r}_{2^{m-2}+\tau-1})}+\omega^{\frac{q}{2}\zeta^1(\mathbf{r}_{3\times 2^{m-2}-\tau})}\right]\\
	&=\left[\omega^{-\frac{q}{2}\eta^0(\mathbf{r}_{2^{m-2}+\tau-1})}+\omega^{-\frac{q}{2}\eta^1(\mathbf{r}_{2^{m-2}+\tau-1})}\right]\\
	&\hspace{0.2cm}-\left[\omega^{\frac{q}{2}\zeta^0(\mathbf{r}_{3\times 2^{m-2}-\tau})}-\omega^{\frac{q}{2}\zeta^1(\mathbf{r}_{3\times 2^{m-2}-\tau})}\right].
	\end{split}
	\end{equation}
	Note that,
	\begin{equation}\label{eq20_t}
	\frac{q}{2}\zeta^1(\mathbf{r}_{3\times 2^{m-2}-\tau})=\frac{q}{2}\left[\zeta^0(\mathbf{r}_{3\times 2^{m-2}-\tau})+r_{3\times 2^{m-2}-\tau,\pi(m-3)}\right],
	\end{equation}
	and
	\begin{equation}\label{eq21_t}
	\frac{q}{2}\eta^0(\mathbf{r}_{2^{m-2}+\tau-1})=\frac{q}{2}\left[\eta^1(\mathbf{r}_{2^{m-2}+\tau-1})+r_{2^{m-2}+\tau-1,\pi(m-3)}\right].
	\end{equation}
	Applying (\ref{eq20_t}) and (\ref{eq21_t}) in (\ref{eq19_t}) we get
	\begin{equation}\label{eq22_t}
	\begin{split}
	&\rho_{\mathbf{a}}(\tau)+\rho_{\mathbf{b}}(\tau)\\
	&=\omega^{-\frac{q}{2}\eta^1(\mathbf{r}_{2^{m-2}+\tau-1})}\left[\omega^{\frac{q}{2}\times r_{2^{m-2}+\tau-1,\pi(m-3)}}+1\right]\\
	&\hspace{0.2cm}-\omega^{\frac{q}{2}\zeta^0(\mathbf{r}_{3\times 2^{m-2}-\tau})}\left[1-\omega^{\frac{q}{2}\times r_{3\times 2^{m-2}-\tau,\pi(m-3)}}\right].
	\end{split}
	\end{equation}
	Again we have the following sub-case
	\begin{itemize}
		\item[1)] Using Lemma \ref{lem4}, for $r_{2^{m-2}+\tau-1,\pi(m-3)}=1$ we have $r_{3\times 2^{m-2}-\tau,\pi(m-3)}=0$. In this case we have from (\ref{eq22_t})
		\begin{equation}
		\rho_{\mathbf{a}}(\tau)+\rho_{\mathbf{b}}(\tau)=0.
		\end{equation}
	\end{itemize}
	For $\tau=2^{m-1}+1$ we have
	\begin{equation}
	\rho_{\mathbf{a}}(\tau)+\rho_{\mathbf{b}}(\tau)=\omega^{-\frac{q}{2}}+1=0.
	\end{equation}
	Hence the ZACZ is $2^{\pi(m-3)}+1$. Now we will check the ZCCZ.

	For calculating ZCCZ we only consider $2^{m-1}-2^{\pi(m-3)}<\tau \leq 2^{m-1}+1$. When $2^{m-1}-2^{\pi(m-3)}<\tau < 2^{m-1}+1$, we have
	\begin{equation}\label{eq25_t}
	\begin{split}
	\rho_{\mathbf{a},\mathbf{b}}(\tau)=&\omega^{-\frac{q}{2}\eta^1(\mathbf{r}_{2^{m-2}+\tau-1})}\\&+\sum_{i=0}^{2^{m-3}-1}\omega^{\frac{q}{2}\zeta^0(\mathbf{r}_{3\times 2^{m-2}-\tau+i})-\frac{q}{2}\eta^1(\mathbf{r}_{2^{m-2}+\tau+i-1})}\\&+\omega^{\frac{q}{2}\zeta^0(\mathbf{r}_{3\times 2^{m-2}-\tau})}
	\end{split}
	\end{equation}
	and
	\begin{equation}\label{eq26_t}
	\begin{split}
	\rho_{\mathbf{b},\mathbf{a}}(\tau)=&\omega^{-\frac{q}{2}\eta^0(\mathbf{r}_{2^{m-2}+\tau-1})}\\&+\sum_{i=0}^{2^{m-3}-1}\omega^{\frac{q}{2}\zeta^1(\mathbf{r}_{3\times 2^{m-2}-\tau+i})-\frac{q}{2}\eta^0(\mathbf{r}_{2^{m-2}+\tau+i-1})}\\&+\omega^{-\frac{q}{2}+\frac{q}{2}\zeta^1(\mathbf{r}_{3\times 2^{m-2}-\tau})}.
	\end{split}
	\end{equation}
	Since from (\ref{eq10_t}) and (\ref{eq11_t}), we have
	\begin{equation}\label{eq27_t}
	\eta^1(\mathbf{r}_{2^{m-2}+\tau+i-1})=\zeta^0(\mathbf{r}_{3\times 2^{m-2}-\tau+i})+\mathbf{1},
	\end{equation}
	and
	\begin{equation}\label{eq28_t}
	\zeta^1(\mathbf{r}_{3\times 2^{m-2}-\tau+i})=\eta^0(\mathbf{r}_{2^{m-2}+\tau+i-1}).
	\end{equation}
	Using (\ref{eq27_t}) and (\ref{eq28_t}) in (\ref{eq25_t}) and (\ref{eq26_t}), respectively, we have
	\begin{equation}\label{eq29_t}
	\begin{split}
	&\rho_{\mathbf{a},\mathbf{b}}(\tau)+\rho_{\mathbf{b},\mathbf{a}}(\tau)\\&=\left[\omega^{-\frac{q}{2}\eta^1(\mathbf{r}_{2^{m-2}+\tau-1})}+\omega^{-\frac{q}{2}\eta^0(\mathbf{r}_{2^{m-2}+\tau-1})}\right]\\&+\sum_{i=0}^{2^{m-3}-1}\left[\omega^{-\frac{q}{2}}+1\right]\\&+\left[\omega^{\frac{q}{2}\zeta^0(\mathbf{r}_{3\times 2^{m-2}-\tau})}+\omega^{-\frac{q}{2}+\frac{q}{2}\zeta^1(\mathbf{r}_{3\times 2^{m-2}-\tau})}\right].
	\end{split}
	\end{equation}
	Further, using (\ref{eq20_t}) and (\ref{eq21_t}) in (\ref{eq29_t}) we get
	\begin{equation}\label{eq30_t}
	\begin{split}
	&\rho_{\mathbf{a},\mathbf{b}}(\tau)+\rho_{\mathbf{b},\mathbf{a}}(\tau)\\&=\omega^{-\frac{q}{2}\eta^1(\mathbf{r}_{2^{m-2}+\tau-1})}\left[1+\omega^{-\frac{q}{2}\times r_{2^{m-2}+\tau-1,\pi(m-3)}}\right]\\&+\omega^{\frac{q}{2}\zeta^0(\mathbf{r}_{3\times 2^{m-2}-\tau})}\left[1-\omega^{\frac{q}{2}\times r_{3\times 2^{m-2}-\tau,\pi(m-3)}}\right].
	\end{split}
	\end{equation}
	We have the following sub-case
	\begin{itemize}
		\item[1)] Using Lemma \ref{lem4}, for $r_{2^{m-2}+\tau-1,\pi(m-3)}=1$ we have $r_{3\times 2^{m-2}-\tau,\pi(m-3)}=0$. In this case we have from (\ref{eq30_t})
		\begin{equation}
		\rho_{\mathbf{a},\mathbf{b}}(\tau)+\rho_{\mathbf{b},\mathbf{a}}(\tau)=0.
		\end{equation}
	\end{itemize}
	For $\tau=2^{m-1}+1$ we have
	\begin{equation}
	\rho_{\mathbf{a},\mathbf{b}}(\tau)+\rho_{\mathbf{b},\mathbf{a}}(\tau)=1+\omega^{-\frac{q}{2}}=0.
	\end{equation}
	Hence the ZCCZ is $2^{\pi(m-3)}+1$. Hence $(\mathbf{a},\mathbf{b})$ is an ($2^{m-1}+2,2^{\pi(m-3)}+1$)- CZCP.
\end{IEEEproof}

Let us take the following example to illustrate the above construction.

\begin{example}\label{ex2}
	Let us consider $q=6$, $m=6$ and $\{\pi(0),\pi(1),\pi(2),\pi(3)\}=\{0,1,2,3\}$. For $d\in \mathbb{Z}_2$, $c=5$. Let the Boolean function $g^d$ be given by
	\begin{equation}
	g^d=3(x_4\bar{x}_5\cdot \zeta^d+\bar{x}_4x_5\cdot \eta^d+\bar{d}x_4x_5)+c,
	\end{equation}
	where
	\begin{equation}
	\zeta^d=x_0x_1+x_1x_2+x_2x_3+dx_3,
	\end{equation}
	and
	\begin{equation}
	\eta^d=\bar{x}_0\bar{x}_1+\bar{x}_1\bar{x}_2+\bar{x}_2\bar{x}_3+\bar{d}\bar{x}_3+d\mathbf{1}.
	\end{equation}
	Since $\pi(3)=3$, the pair $(\mathbf{a},\mathbf{b})= \left(\Psi_{15}(g^0),\Psi_{15}(g^1)\right)$ is constructed as follows:
	\begin{equation}
	\begin{split}
	\mathbf{a}&=\omega_6^{[5,5,5,5,2,5,5,2,5,5,5,5,2,2,2,5,2,5,2,5     ,5,5,2,2,2,5,2,5,5,2,5,5,5,2]},\\
	\mathbf{b}&=\omega_6^{[5,5,5,5,2,5,5,2,5,2,2,2,5,5,5, 2,5,5,2,5,5     ,5,2,2,2,2,5,2,2,5, 2,2,2 ,5
		]}.\\
	\end{split}
	\end{equation}
	Then,
	\begin{equation}
	\{|\rho_{\mathbf{a}}(\tau)+\rho_{\mathbf{b}}(\tau)|\}_{\tau=0}^{33}=(68,\mathbf{0}_{16},\mathbf{4}_{8},\mathbf{0}_{9}).
	\end{equation}
	And,
	\begin{equation}
	\begin{split}
	&\{|\rho_{\mathbf{a},\mathbf{b}}(\tau)+\rho_{\mathbf{b},\mathbf{a}}(\tau)|\}_{\tau=0}^{33}=(0,\mathbf{4}_{9},8,12,0,4,8,4,0,\\&\hspace{4cm}4,8,4,0,12,8,4,\mathbf{0}_{10}).
	\end{split}
	\end{equation}
	Hence, $(\mathbf{a},\mathbf{b})$ is a $(34,9)$- CZCP. \end{example}

In the next section we propose CZCPs with more new lengths using insertion method.
	
	\section{Construction of CZCPs Through Insertion Method}
	In this section, we propose constructions of CZCPs of lengths of the form $2^{\alpha} 10^\beta 26^\gamma+2 ~(\alpha\geq1)$, $10^\beta+2$, $26^\gamma+2$ and $10^\beta 26^\gamma+2$.


\begin{theorem}\label{th1_insertion}
	Let $(\mathbf{a},\mathbf{b})$ be a binary GCP of length $N=2^\alpha10^\beta26^\gamma ~(\alpha\geq1)$ constructed via Result \ref{r2}, and $(\mathbf{c},\mathbf{d})$ be one of its complementary mate. Also let $\mathbf{e}=(\mathbf{a}||\mathbf{c})$, $\mathbf{f}=(\mathbf{b}||\mathbf{d})$, $\mathbf{g}=\mathcal{I}_s(\mathbf{e},0,\{x_0,y_0\})$, and $\mathbf{h}=\mathcal{I}_s(\mathbf{f},0,\{x_1,y_1\})$ where $x_0,~y_0,~x_1,~y_1\in \mathbb{U}_q$. Then $(\mathbf{g},\mathbf{h})$ is $(2N+2,N/2+1)$- CZCP if the following conditions hold.
	\begin{equation}\label{eq10}
		\begin{split}
		& x_0-y^*_1=0,~x_1+y^*_0=0\\
		& x_0=x_1,~ y^*_0=-y^*_1.
		\end{split}
	\end{equation}
\end{theorem}
\begin{IEEEproof}
	From \cite{Avik_tit} we know that $(\mathbf{a},\mathbf{b})$ is a GCP having the property
	\begin{equation}\label{eq11}
	\begin{split}
	a_i=b_i & \text{ for }0\leq i <N/2,\\
	a_i=-b_i & \text{ for }N/2\leq i <N.
	\end{split}
	\end{equation}
 Let us define two sets $\mathcal{A}=\{0,2\}$ and $\mathcal{B}=\{1,3\}$. Also let $L=N/2$. Define $\sigma$ as follows:
\begin{equation}
	\sigma(i)=\begin{cases}
		1 &\text{ if }i\in \mathcal{A}\\
		-1 &\text{ if }i\in \mathcal{B}
	\end{cases}
\end{equation}

	Consider $(\mathbf{c},\mathbf{d})$ as one of the complementary mates of $(\mathbf{a},\mathbf{b})$, then $(\mathbf{c},\mathbf{d})$ will also have the same property as in (\ref{eq11}). Consequently, $\mathbf{e}$ and $\mathbf{f}$ are sequences of length $2N$, having the following structural property
\begin{equation}\label{eq13}
f_i=\sigma\left(\lfloor i/L \rfloor\right) e_i.
\end{equation}
Also within sequence $\mathbf{e}$, we have the following property
\begin{equation}\label{eq13.1}
e_i=\begin{cases}
e_{2N-1-i} & \text{ if }0\leq i <L\\
-e_{2N-1-i} & \text{ if }L\leq i <N
\end{cases}.
\end{equation}
Define $\delta_{\tau,L}$ as
\begin{equation}
	\delta_{\tau,L}=\begin{cases}
		0 &\text{if } \tau \text{ mod }L\equiv 0 \text{ or }\tau \geq 3N/2,\\
		1 &\text{otherwise }
	\end{cases}
\end{equation}
Also, let $\mathbf{e}|^M_n$ denote a subsequence of $\mathbf{e}$, containing $n$ consecutive elements of $\mathbf{e}$, starting from the index $M L$. First, let us calculate the autocorrelation to check condition C1 of (\ref{con_eq}). For $i L\leq \tau < j L$, where $0\leq i <4$, $j=i+1$, we have
\begin{equation}\label{eq14}
	\begin{split}
		\rho_{\mathbf{g}}(\tau)=&x_0e_{\tau-1}+\sum_{k=0}^{4-j}   \rho_{\mathbf{e}|^{k}_{L},\mathbf{e}|^{k+i}_{L}}(\tau-i L)\\&+ \delta_{\tau,L}\sum_{k=0}^{3-j} \rho_{\overleftarrow{\mathbf{e}|^{k}_{L}},\overleftarrow{\mathbf{e}|^{k+j}_{L}}}(j L-\tau)+e_{2N-\tau}y^*_0.
	\end{split}
\end{equation}
For $\tau=4L$, we have
\begin{equation}\label{eq15.1}
\rho_{\mathbf{g}}(\tau)=x_0e_{2N-1}+e_0y^*_0.
\end{equation}
For $\tau=4L+1$, we have
\begin{equation}\label{eq15}
\rho_{\mathbf{g}}(\tau)=x_0y^*_0.
\end{equation}

Similarly, we calculate the autocorrelation of sequence $\mathbf{h}$. For $i L\leq \tau < j L$, where $0\leq i <4$, $j=i+1$, we have
\begin{equation}\label{eq_auto}
\begin{split}
\rho_{\mathbf{h}}(\tau)=&x_1f_{\tau-1}+\sum_{k=0}^{4-j} \rho_{\mathbf{f}|^{k}_{L},\mathbf{f}|^{k+i}_{L}}(\tau-i L)\\&+ \delta_{\tau,L}\sum_{k=0}^{3-j} \rho_{\overleftarrow{\mathbf{f}|^{k}_{L}},\overleftarrow{\mathbf{f}|^{k+j}_{L}}}(j L-\tau)+f_{2N-\tau}y^*_1.
\end{split}
\end{equation}
For $\tau=4L$, we have
\begin{equation}\label{eq19.2}
\rho_{\mathbf{h}}(\tau)=x_1f_{2N-1}+f_0y^*_1.
\end{equation}
For $\tau=4L+1$, we have
\begin{equation}\label{eq18}
\rho_{\mathbf{h}}(\tau)=x_1y^*_1.
\end{equation}

 Recall that ($\mathbf{e}$,$\mathbf{f}$) is a GCP, using (\ref{eq13}) and (\ref{eq13.1}) we get
\begin{equation}\label{eq20}
\begin{split}
	&\rho_{\mathbf{g}}(\tau)+\rho_{\mathbf{h}}(\tau)=\\&\begin{cases}
	(x_0+y^*_0+x_1-y^*_1)e_{\tau-1}& \text{ if }0<\tau \leq N/2,\\
	(x_0-y^*_0-x_1-y^*_1)e_{\tau-1}	& \text{ if }N/2<\tau \leq N,\\
	(-x_0+y^*_0+x_1+y^*_1)e_{2N-\tau}& \text{ if }3N/2<\tau < 2N,\\
	(x_0+y^*_0-x_1+y^*_1)e_{2N-1}& \text{ if }\tau=2N,\\
x_0y^*_0+x_1y^*_1 & \text{ if }\tau=2N+1.
\end{cases}
\end{split}
\end{equation}
Therefore we can get a ZACZ of $(N/2+1)$ for the sequence pair ($\mathbf{g}$, $\mathbf{h}$) if the given conditions hold.

To check the condition C2, we calculate the cross-correlation of $\mathbf{g}$ and $\mathbf{h}$. For $i L\leq \tau < j L$, where $0\leq i <4$, $j=i+1$, we have
\begin{equation}
\begin{split}
&\rho_{\mathbf{g},\mathbf{h}}(\tau)=\\&\sigma(i)\cdot x_0e_{\tau-1}+\sum_{k=0}^{4-j}  \sigma(k+i)\cdot   \rho_{\mathbf{e}|^{k}_{L},\mathbf{e}|^{k+i}_{L}}(\tau-i L)\\&+\delta_{\tau,L} \sum_{k=0}^{3-j} \sigma(k+j)\cdot \rho_{\overleftarrow{\mathbf{e}|^{k}_{L}},\overleftarrow{\mathbf{e}|^{k+j}_{L}}}(j L-\tau)+\sigma(i)\cdot e_{2N-\tau}y^*_1.
\end{split}
\end{equation}
For $\tau=4L$, we have
\begin{equation}\label{eq24}
\begin{split}
\rho_{\mathbf{g},\mathbf{h}}(\tau)&=x_0f_{2N-1}+e_0y^*_1\\
&=-x_0e_{2N-1}+e_0y^*_1\\
&=(-x_0+y^*_1)e_0.
\end{split}
\end{equation}
For $\tau=4L+1$, we have
\begin{equation}
\rho_{\mathbf{g},\mathbf{h}}(\tau)=x_0y^*_1.
\end{equation}
Similarly, for $i L\leq \tau < j L$, where $0\leq i <4$, $j=i+1$, we have
\begin{equation}
\begin{split}
&\rho_{\mathbf{h},\mathbf{g}}(\tau)=\\& x_1e_{\tau-1}+\sum_{k=0}^{4-j}  \sigma(k)\cdot   \rho_{\mathbf{e}|^{k}_{L},\mathbf{e}|^{k+i}_{L}}(\tau-i L)\\&+\delta_{\tau,L} \sum_{k=0}^{3-j} \sigma(k)\cdot \rho_{\overleftarrow{\mathbf{e}|^{k}_{L}},\overleftarrow{\mathbf{e}|^{k+j}_{L}}}(j L-\tau)+ e_{2N-\tau}y^*_0.
\end{split}
\end{equation}
For $\tau=4L$, we have
\begin{equation}\label{eq27}
\begin{split}
\rho_{\mathbf{h},\mathbf{g}}(\tau)&=x_1e_{2N-1}+e_0y^*_0\\
&=(x_1+y^*_0)e_0.
\end{split}
\end{equation}
For $\tau=4L+1$, we have
\begin{equation}
\rho_{\mathbf{h},\mathbf{g}}(\tau)=x_1y^*_0.
\end{equation}

Then for $1\leq \tau < 2N$ $\rho_{\mathbf{g},\mathbf{h}}(\tau)+\rho_{\mathbf{h},\mathbf{g}}(\tau)$ is given in (\ref{eq29}).

\begin{figure*}
	\begin{equation}\label{eq29}
	\begin{split}
		\rho_{\mathbf{g},\mathbf{h}}(\tau)+\rho_{\mathbf{h},\mathbf{g}}(\tau)=&
	(\sigma(i)\cdot x_0+x_1)e_{\tau-1}+\sum_{k=0}^{4-j}  (\sigma(k+i)+\sigma(k))\cdot   \rho_{\mathbf{e}|^{k}_{L},\mathbf{e}|^{k+i}_{L}}(\tau-i L)\\&+\delta_{\tau,L} \sum_{k=0}^{3-j} (\sigma(k+i)+\sigma(k))\cdot \rho_{\overleftarrow{\mathbf{e}|^{k}_{L}},\overleftarrow{\mathbf{e}|^{k+j}_{L}}}(j L-\tau)+ e_{2N-\tau}(\sigma(i)\cdot y^*_1+y^*_0).
	\end{split}
	\end{equation}
\end{figure*}

For rest of the cases, we have
\begin{equation}\label{eq30}
\begin{split}
	&\rho_{\mathbf{g},\mathbf{h}}(\tau)+\rho_{\mathbf{h},\mathbf{g}}(\tau)=\\&\begin{cases}
(-x_0+y^*_1+x_1+y^*_0)e_0 & \text{ if }\tau=2N,\\
x_0y^*_1+x_1y^*_0 & \text{ if }\tau=2N+1.
\end{cases}
\end{split}
\end{equation}

Since according to the setup $\sigma(i)=-1$ when $3N/2 \leq \tau <2N$, $\rho_{\mathbf{g},\mathbf{h}}(\tau)+\rho_{\mathbf{h},\mathbf{g}}(\tau)=0$ if $(-x_0+x_1-y^*_1+y^*_0)=0$. Therefore, from (\ref{eq30}) and the above explanation, we conclude that the ZCCZ of ($\mathbf{g}$, $\mathbf{h}$) is $(N/2+2)$.

Hence, we conclude that ($\mathbf{g}$, $\mathbf{h}$) is a ($2N+2$, $N/2+1$)- CZCP.
\end{IEEEproof}

In the following example we will illustrate the proposed construction step by step.
\begin{example}\label{ex1}
	Step 1: Let $(\mathbf{a},\mathbf{b})$ be a GCP of length $8$, constructed via Result \ref{r2} as follows:
\begin{equation}\label{examp2_10}
	\left( \begin{matrix}
		\mathbf{a} \\
		\mathbf{b}
	\end{matrix} \right)= \left ( \begin{matrix}
		\textcolor{blue}{+++-}++-+\\
		\textcolor{blue}{+++-}--+-
	\end{matrix} \right)
\end{equation}
Step 2: Let $(\mathbf{c},\mathbf{d})$ be a complementary mate of $(\mathbf{a},\mathbf{b})$. Therefore,
\begin{equation}
	\left( \begin{matrix}
		\mathbf{c} \\
		\mathbf{d}
	\end{matrix} \right)= \left ( \begin{matrix}
		\textcolor{blue}{+-++}-+++\\
		\textcolor{blue}{+-++}+---
	\end{matrix} \right)
\end{equation}	
Step 3: Define $\mathbf{e}=(\mathbf{a}||\mathbf{c})$ and $\mathbf{f}=(\mathbf{b}||\mathbf{d})$. Therefore,
\begin{equation}
	\left( \begin{matrix}
		\mathbf{e} \\
		\mathbf{f}
	\end{matrix} \right)= \left ( \begin{matrix}
		\textcolor{blue}{+-++}-+++\textcolor{blue}{+-++}-+++\\
		\textcolor{blue}{+-++}+---\textcolor{blue}{+-++}+---
	\end{matrix} \right)
\end{equation}	
Step 4: Set $\mathbf{g}=\mathcal{I}_s(\mathbf{e},0,\{x_0,y_0\})$, and $\mathbf{h}=\mathcal{I}_s(\mathbf{f},0,\{x_1,y_1\})$. Let $x_0=1,~x_1=1,~y_0=-1,$ and $y_1=1,$Therefore,
\begin{equation}
	\left( \begin{matrix}
		\mathbf{g} \\
		\mathbf{h}
	\end{matrix} \right)= \left ( \begin{matrix}
		\textcolor{red}{+}\textcolor{blue}{+-++}-+++\textcolor{blue}{+-++}-+++\textcolor{red}{-}\\
		\textcolor{red}{+}\textcolor{blue}{+-++}+---\textcolor{blue}{+-++}+---\textcolor{red}{+}
	\end{matrix} \right)
\end{equation}
The pair $(\mathbf{g},\mathbf{h})$ is a length $(18,5)$- CZCP, because
\begin{equation}
	\begin{array}{ccl}
		| \rho_{\mathbf{g}}(\tau)+\rho_{\mathbf{h}}(\tau) |_{\tau=0}^{17} & = & (36,\mathbf{0}_{8},\mathbf{4}_{4},\mathbf{0}_{5}).
	\end{array}
\end{equation}
and
\begin{equation}
\begin{array}{ccl}
| \rho_{\mathbf{g},\mathbf{h}}(\tau)+\rho_{\mathbf{h},\mathbf{g}}(\tau) |_{\tau=0}^{17} & = & (0,\mathbf{4}_{5},8,4,0,4,8,4,\mathbf{0}_{6}).
\end{array}
\end{equation}		
\end{example}

\begin{figure}
	\includegraphics[width=\columnwidth,draft=false]{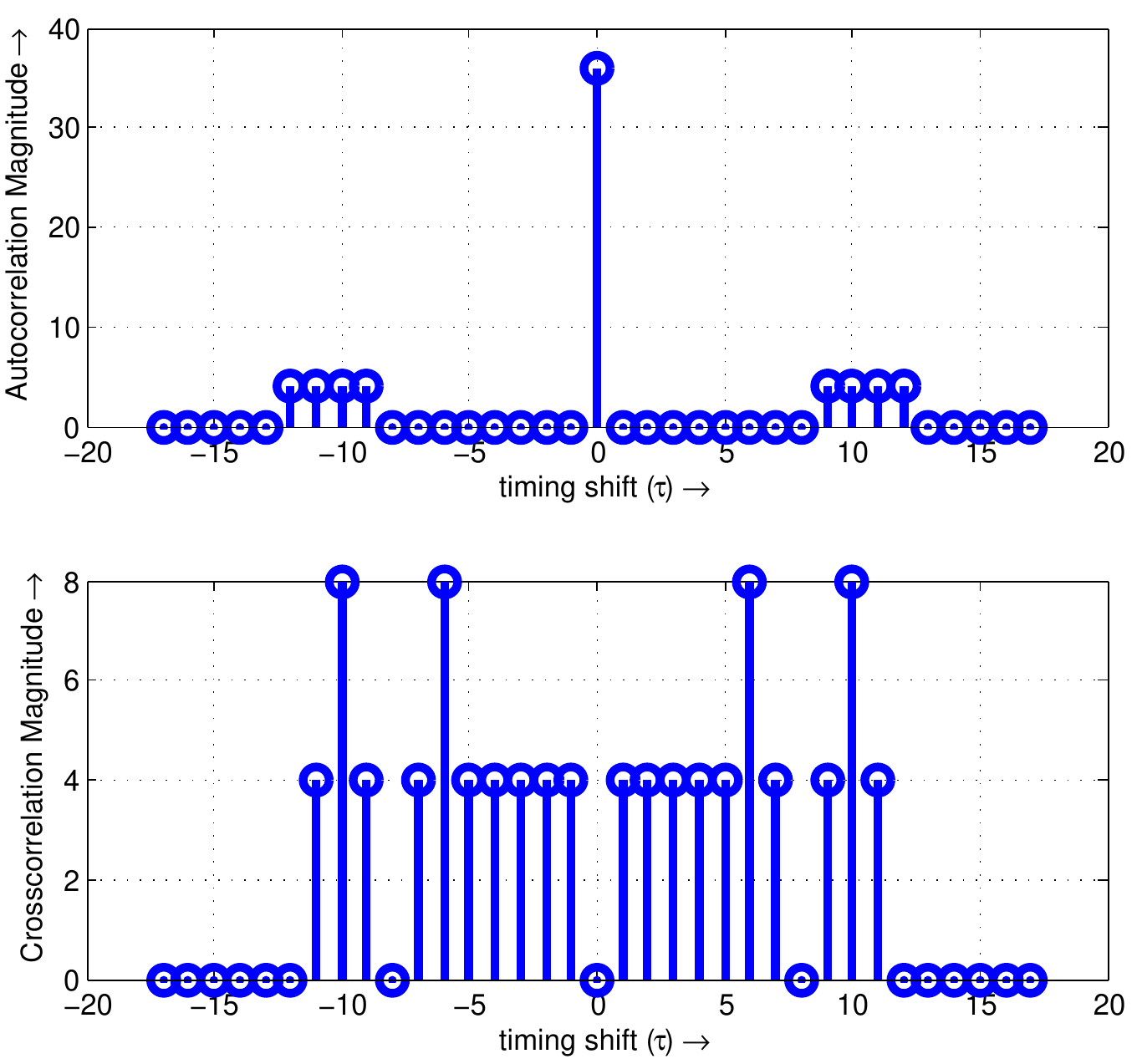}
	\caption{A glimpse of correlation magnitudes of the $(18,5)$- CZCP in \textit{Example \ref{ex1}}.}\label{label-a}
\end{figure}

\begin{remark}
Theorem \ref{th1_insertion} can include the results of Theorem \ref{th1_gbf} with $q = 2$.
\end{remark}

We give the following theorem without proof, since the proof is similar to Theorem \ref{th1_insertion}.

\renewcommand{\arraystretch}{0.6}
\begin{table}
	\small
	\centering
	\tabcolsep=0.11cm
	\caption{CZCPs constructed in Theorem \ref{th_3}\label{czcp_insertion}.}
	\resizebox{\columnwidth}{!}{
		\begin{tabular}{|c||c||c|}
			\hline
			$N$ & \makecell{Parameters of the\\ proposed CZCPs} & Remarks   \\ \hline \hline
			$10^\beta$  & $(2N+2,4N/10+1)$ & \makecell{GCP is constructed\\ using Result \ref{r3}.}  \\  \hline
			$26^\gamma$  & $(2N+2,12N/26+1)$ & \makecell{GCP is constructed\\ using Result \ref{r3}.}  \\  \hline
			$10^\beta26^\gamma$  & $(2N+2,12N/26+1)$ & \makecell{GCP is constructed\\ using Result \ref{r4}.}  \\  \hline
		\end{tabular}
	}
\end{table}

\begin{theorem}\label{th_3}
	Let $(\mathbf{a},\mathbf{b})$ be a binary GCP of length $N$ constructed via Turyn's method as given in Table \ref{czcp_insertion}, and $(\mathbf{c},\mathbf{d})$ be one of its complementary mate. Also let $\mathbf{e}=(\mathbf{a}||\mathbf{c})$, $\mathbf{f}=(\mathbf{b}||\mathbf{d})$, $\mathbf{g}=\mathcal{I}_s(\mathbf{e},0,\{x_0,y_0\})$, and $\mathbf{h}=\mathcal{I}_s(\mathbf{f},0,\{x_1,y_1\})$ where $x_0,~y_0,~x_1,~y_1\in \mathbb{U}_q$. Then $(\mathbf{g},\mathbf{h})$ is a CZCP, as given in Table \ref{czcp_insertion} if the following conditions hold.
	\begin{equation}\label{eq110}
	\begin{split}
	& x_0-y^*_1=0,~x_1+y^*_0=0\\
	& x_0=x_1,~ y^*_0=-y^*_1.
	\end{split}
	\end{equation}
\end{theorem}

\section{Construction of CZCPs Using Barker Sequences}

Before going to the construction, let us recall the binary Barker sequences in Table \ref{barker_tab}. Since the sequences are specific, we list down their autocorrelations as well.

\begin{table}
	\small
	\centering
	\tabcolsep=0.11cm
	\caption{\cite{fan_book} List of binary Barker sequences. \label{barker_tab}}
	\resizebox{\columnwidth}{!}{
		
		\begin{tabular}{|c|c|c|}
			\hline
			$N$ & Barker sequence/s ($\mathbf{a}$) & $ \left\lbrace  \rho_{\mathbf{a}}(\tau) \right\rbrace_{\tau=0}^{N-1}$   \\ \hline \hline
			$2$  & $+-$, $++$ & $\{2,-1\}$, $\{2,1\}$ \\  \hline
			$3$ & $++-$ & $\{3,0,-1\}$ \\  \hline
			$4$ & $+-++$, $+---$ & $\{4,-1,0,1\}$, $\{4,1,0,-1\}$ \\  \hline
			$5$ & $+++-+$& $\{5,0,1,0,1\}$ \\ \hline
			$7$ & $+++--+-$& $\{7,0,-1,0,-1,0,-1\}$ \\ \hline
			$11$ & $+++---+--+-$& $\{11,0,-1,0,-1,0,-1,0,-1\}$ \\ \hline
			$13$ & $+++++--++-+-+$& $\{13,0,1,0,1,0,1,0,1,0,1,0,1\}$ \\ \hline
		\end{tabular}
	}
\end{table}

Using their correlation properties given in Table \ref{barker_tab}, we have the following theorem.

\begin{theorem}\label{th6}
	Let $\mathbf{a}$ and $\mathbf{b}$ be binary Barker sequences of lengths $M$ and $N$, respectively, and $M\leq N$. Then the sequence pair ($\mathbf{c}$,$\mathbf{d}$) given by
	\begin{equation}\label{eq75}
	\begin{split}
	\mathbf{c}&=\mathbf{a}||\mathbf{b};\\
	\mathbf{d}&=\mathbf{a}||-\mathbf{b};\\
	\end{split}	
	\end{equation}
	forms an ($M+N,M$)- CZCP, if $\rho_{\mathbf{a}}(\tau)=-\rho_{\mathbf{b}}(\tau)$ for $0<\tau<M$ and $\rho_{\mathbf{b}}(M)=0$ when $M<N$.
\end{theorem}

\begin{IEEEproof}
	As per the condition given in (\ref{eq75}), we have for $0<\tau<M$,
	\begin{equation}\label{eq76}
		\rho_{\mathbf{c}}(\tau)+\rho_{\mathbf{d}}(\tau)=0,
	\end{equation}
	since, $\rho_{\mathbf{a}}(\tau)=-\rho_{\mathbf{b}}(\tau)$ for $0<\tau<M$.
	
	For $M\leq\tau<N$, we have
	\begin{equation}\label{eq77}
		\rho_{\mathbf{c}}(\tau)+\rho_{\mathbf{d}}(\tau)=2\rho_{\mathbf{b}}(\tau).
	\end{equation}
	
	For $N\leq\tau<M+N$, we have
	\begin{equation}
	\rho_{\mathbf{c}}(\tau)+\rho_{\mathbf{d}}(\tau)=0.
	\end{equation}
	
	Now, to analyse the cross-correlation, we have for $0<\tau <M$,
	\begin{equation}
	\begin{split}
		\rho_{\mathbf{c},\mathbf{d}}(\tau)&=\rho_{\mathbf{d}}(\tau)-2\rho_{\mathbf{b}}(\tau);\\
		\rho_{\mathbf{d},\mathbf{c}}(\tau)&=\rho_{\mathbf{c}}(\tau)-2\rho_{\mathbf{b}}(\tau).
	\end{split}
	\end{equation}
	Therefore, from (\ref{eq76}) we have for $0<\tau <M$,
	\begin{equation}
		\rho_{\mathbf{c},\mathbf{d}}(\tau)+\rho_{\mathbf{d},\mathbf{c}}(\tau)=-4\rho_{\mathbf{b}}(\tau).
	\end{equation}
	Similarly, we have for $M\leq\tau <N$,
	\begin{equation}
	\rho_{\mathbf{c},\mathbf{d}}(\tau)+\rho_{\mathbf{d},\mathbf{c}}(\tau)=-2\rho_{\mathbf{b}}(\tau).
	\end{equation}
	And, for $N\leq\tau <M+N$, we have
	\begin{equation}
	\rho_{\mathbf{c},\mathbf{d}}(\tau)+\rho_{\mathbf{d},\mathbf{c}}(\tau)=0.
	\end{equation}
	Hence the theorem is proved.
\end{IEEEproof}

\begin{remark}
	According to Theorem \ref{th6} and Table \ref{barker_tab}, we can obtain an optimal binary $(12,5)$- CZCP by taking $M=5$ and $N=7$. When $M=11$ and $N=13$, using Table \ref{barker_tab}, we get an optimal binary $(24,11)$- CZCP. Note that these CZCPs are different from that of the ``best possible" CZCPs given in \cite[Table I]{zilong_ccp}, obtained through computer search.
\end{remark}

\begin{property}\label{pro1}
	Let ($\mathbf{c}$,$\mathbf{d}$) be a binary $(N,Z)$- CZCP, then
	\begin{equation}
	c_i=d_i, \text{ and } c_{N-1-i}=-d_{N-1-i},
	\end{equation}
	for all $i\in \{0,1,\cdots,Z-1\}$.
\end{property}
\begin{IEEEproof}
	If $a,~b$ can only take the values $+1$ or $-1$, then we know that
	\begin{equation}\label{quad86}
	ab\equiv a+b-1 \pmod 4.
	\end{equation}
	From C2 of (\ref{con_eq}), we get for $Z\leq \tau <N$
	\begin{equation}
	\begin{split}
	c_0d_{N-1}+d_0c_{N-1}&=0,\\
	c_0d_{N-2}+c_1d_{N-1}+d_0c_{N-2}+d_1c_{N-1}&=0,\\
	&\vdots\\
	c_0d_{N-Z}+\dots+c_{Z-1}d_{N-1}\hspace{3cm}\\+d_0c_{N-Z}+\dots+d_{Z-1}c_{N-1}&=0.
	\end{split}
	\end{equation}
	Using (\ref{quad86}) to reduce the above equations, we get
	\begin{equation}
	c_i+d_i+c_{N-1-i}+d_{N-1-i}=2 \pmod 4,
	\end{equation}
	which is equivalent to
	\begin{equation}
	c_ic_{N-1-i}+d_id_{N-1-i}=0.
	\end{equation}
	Equivalently, we can conclude that $c_i=d_i$ and $c_{N-1-i}=-d_{N-1-i}$.
\end{IEEEproof}

\begin{theorem}\label{th_new7}
	Any binary GCP $(\mathbf{a},\mathbf{b})$ of length $N$ is also a CZCP.
\end{theorem}
\begin{IEEEproof}
	We just need to check C2 of (\ref{con_eq}) for $\tau=N-1$. We already know that
	\begin{equation}\label{eq83}
		a_0a_{N-1}+b_0b_{N-1}=0.
	\end{equation}
	Using (\ref{quad86}) we can write (\ref{eq83}) as
	\begin{equation}
	\begin{split}
			a_0+a_{N-1}-1+b_0+b_{N-1}-1&\equiv 0 \pmod 4,\\
		    \text{or, }a_0+b_{N-1}-1+b_0+a_{N-1}-1&\equiv 0 \pmod 4.
	\end{split}
	\end{equation}
	Therefore, we can conclude that
	\begin{equation}
	a_0b_{N-1}+b_0a_{N-1}=0.
	\end{equation}
	This completes the proof.
\end{IEEEproof}

In the next theorem we will enlarge the length of the CZCPs generated by Theorem \ref{th6} with the help of GCPs and utilizing Turyn's method.


\begin{theorem}\label{th7}
	Let $\mathcal{A}=$($\mathbf{a}$,$\mathbf{b}$) be a GCP of length $M$ and $\mathcal{B}=$($\mathbf{c}$,$\mathbf{d}$) be a $(N,Z)$- CZCP. Also, let \begin{equation}
		(\mathbf{e},\mathbf{f})=Turyn(\mathcal{A},\mathcal{B}).
	\end{equation}
	Then $(\mathbf{e},\mathbf{f})$ is an $(NM,ZM)$- CZCP.
\end{theorem}
\begin{IEEEproof}
See Appendix.
\end{IEEEproof}

All the parameters of the CZCPs which can be generated through systematic construction, including this paper, are listed in Table \ref{list}.

\begin{table*}
	\small
	\centering
	\tabcolsep=0.11cm
	\caption{Parameters of the CZCPs. \label{list}}
	
	\begin{tabular}{|c|c|c|c|c|}
		\hline
		Ref.&  Length of the CZCP & $Z$ &$CZC_{ratio}$ &Remarks   \\ \hline \hline
		\cite{zilong_ccp}&  $2^\alpha$ & $2^{\alpha-1}$ &$1$& Optimal\\ \hline
		\cite{zilong_ccp}&  \makecell{$2^{\alpha+1}10^\beta 26^\gamma$\\($\alpha\geq 1$)} & $2^{\alpha}10^\beta 26^\gamma$ &$1$& Optimal\\ \hline
		Th. \ref{th1_gbf}&  \makecell{$2^{m-1}+2$\\($m\geq 4$)} & $2^{m-3}+1$ & $\approx \frac{1}{2}$&Not optimal \\ \hline
		Th. \ref{th1_insertion}&  \makecell{$2^{\alpha+1}10^\beta 26^\gamma+2$\\($\alpha\geq 1$)} & $2^{\alpha-1}10^\beta 26^\gamma+1$ &$\approx \frac{1}{2}$& Not optimal \\ \hline
		Th. \ref{th_3}&  \makecell{$2N+2$\\ ($N=10^\beta$)} & $4N/10+1$&$\approx \frac{2}{5}$& Not optimal \\ \hline
		Th. \ref{th_3}&  \makecell{$2N+2$\\ ($N=26^\gamma$)} & $12N/26+1$&$\approx \frac{6}{13}$& Not optimal \\ \hline
		Th. \ref{th_3}&  \makecell{$2N+2$\\ ($N=10^\beta26^\gamma$)} & $12N/26+1$&$\approx \frac{6}{13}$& Not optimal \\ \hline
		Th. \ref{th6}&  $12$ & $5$&$1$& Optimal \\ \hline
		Th. \ref{th6}&  $24$ & $11$&$1$& Optimal \\ \hline
		Th. \ref{th7}&  \makecell{$12N$\\($N=2^\alpha 10^\beta 26^\gamma$)} & $5N$&$\approx \frac{5}{6}$&Large $CZC_{ratio}$ \\ \hline
		Th. \ref{th7}&  \makecell{$24N$\\($N=2^\alpha 10^\beta 26^\gamma$)} & $11N$&$\approx \frac{11}{12}$& Large $CZC_{ratio}$ \\ \hline
	\end{tabular}
	
\end{table*}

\section{Performance Analysis of the Proposed CZCPs to Design Training Sequences for SM Systems}

In this section, we will analyse the performance of the proposed CZCPs in designing training sequences for SM systems, based on the framework proposed in \cite{zilong_ccp}.

Let us consider a generic training-based single carrier MIMO transmission structure with $N_t$ transmit antennas and $N_r$ receive antennas. Assume that each of the antennas has $(\lambda+1)$- multipath in an additive white Gaussian noise (AWGN) channel having zero mean and variance $\sigma^2/2$. Consider that each transmit antenna will send $Q=J\theta$ non-zero entries over $J$ sub-blocks (where $J\geq 2$ and $\theta$ is a natural number) so that they can work in a co-operative way to enable the resultant training matrix to meet the optimal condition \cite{zilong_ccp}. Applying the least-squares (LS) channel estimator (unbiased) the normalized mean square error (MSE) can be derived as \cite{zilong_ccp}
\begin{equation}
	MSE=\frac{\sigma^2}{N_t\lambda+N_t}Tr((X^HX)^{-1}),
\end{equation}
where $X$ is a $N_tJ\theta \times N_t(\lambda+1)$ training matrix. Since the non-zero entries of the training sequences have ideal magnitude of $1$, so in this case, the minimum MSE is given by $\sigma^2/Q$.

Let $N_t=4$ and $N_r=1$ and $J\in\{2,6,18\}$. Consider an optimal $(12,5)$- CZCP constructed through Theorem \ref{th6}. Let us design a training matrix for SM system, considering $5$ multi-path. We will take a random sequence for comparison.
\begin{figure}
	\centering
	\includegraphics[width=\columnwidth,draft=false]{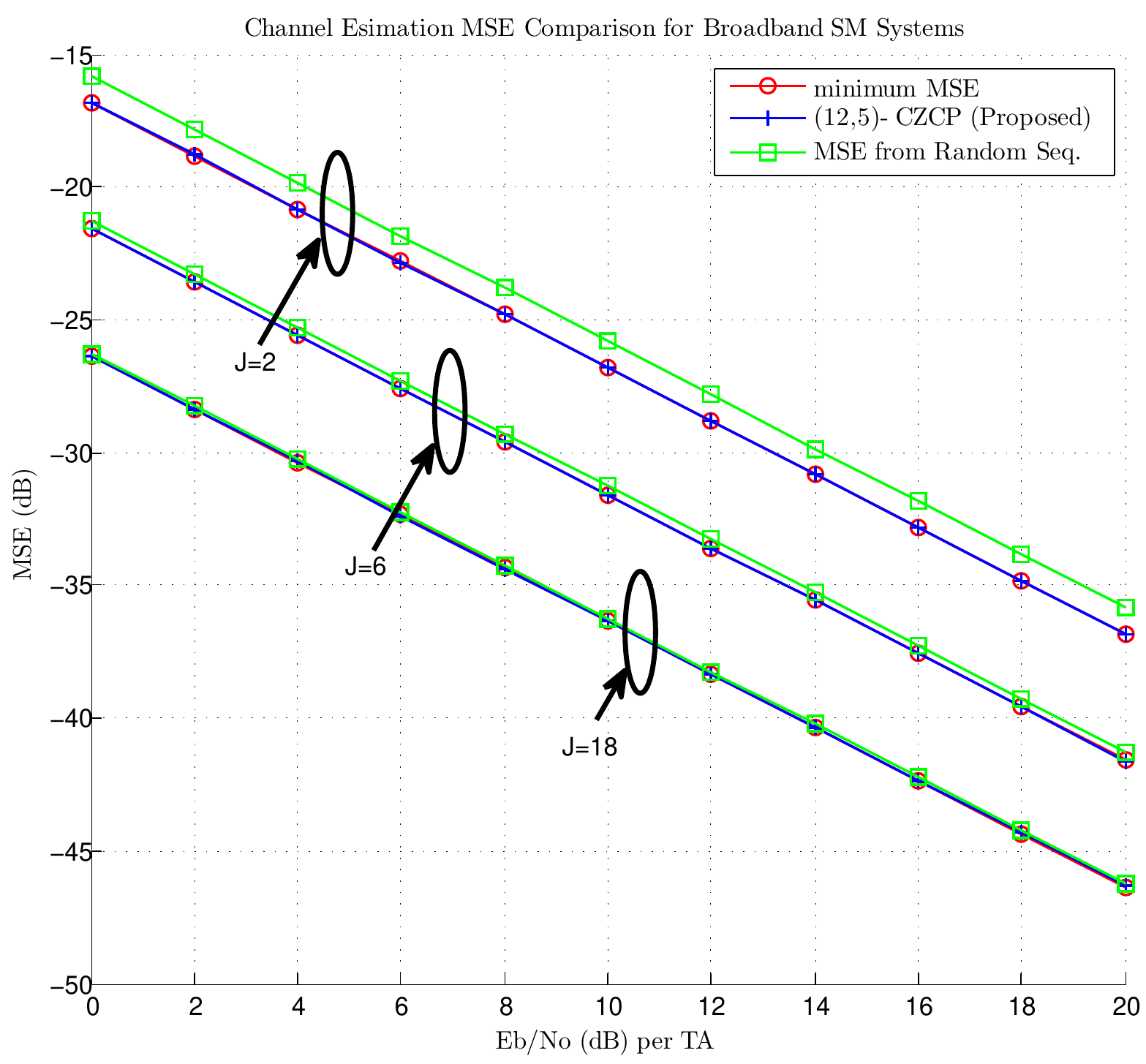}
	\caption{MSE comparison, No. of multi-paths 5, $(12,5)$- CZCP. \label{j2j6}}
\end{figure}
Fig. \ref{j2j6} shows that when the number of multi-paths is $5$ (or less), $(12,5)$- CZCP constructed through Theorem \ref{th6} can be used to design SM training matrix which attains the MSE lower bound.

In a similar set-up, we consider an $(18,5)$- CZCP derived in Example \ref{ex1}.
\begin{figure}
	\centering
	\includegraphics[width=\columnwidth,draft=false]{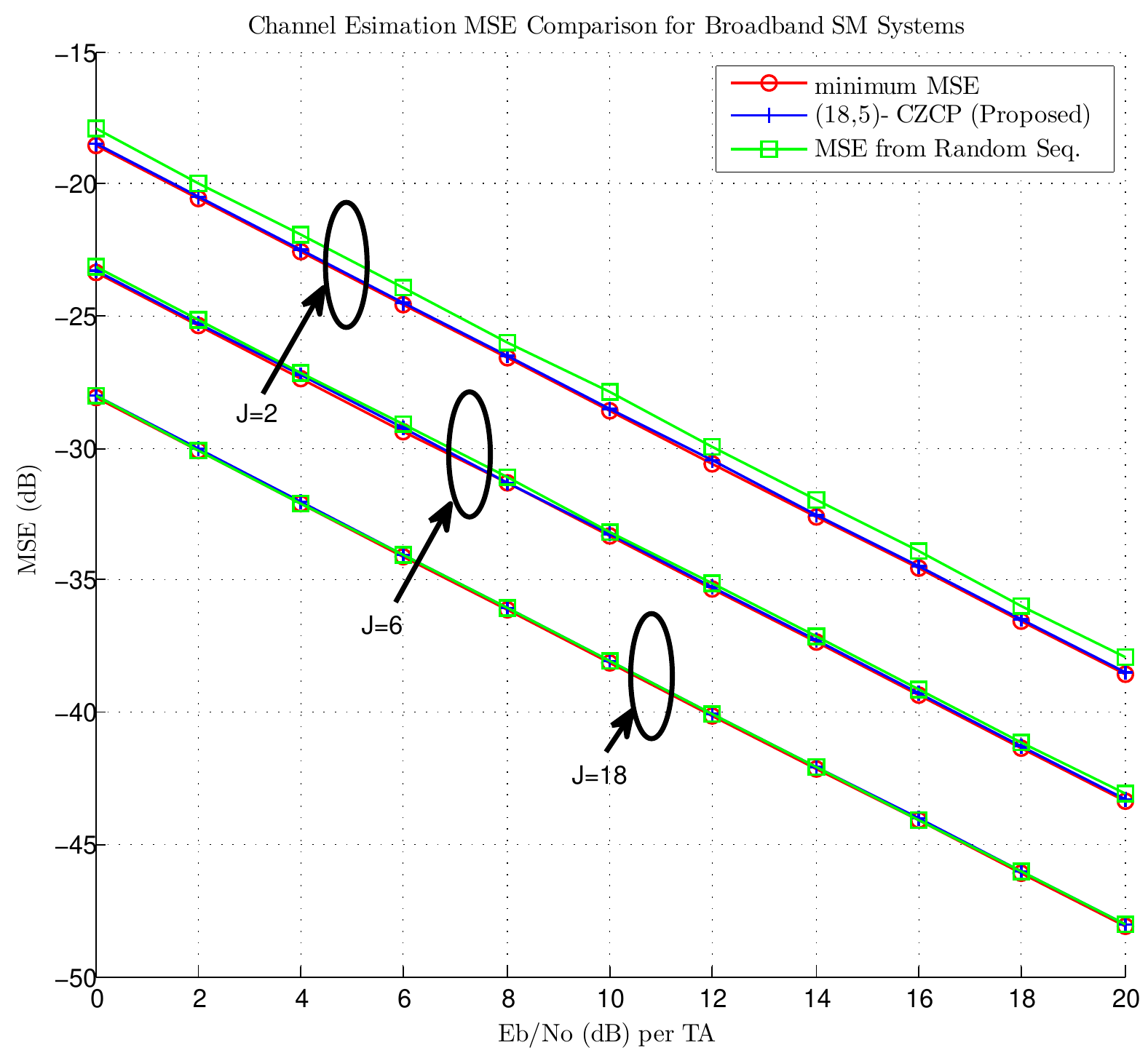}
	\caption{MSE comparison, No. of multi-paths 5, $(18,5)$- CZCP. \label{j2j6len18}}
\end{figure}
Fig. \ref{j2j6len18} shows that when the number of multi-paths is $5$ (or less), $(18,5)$- CZCP constructed in Example \ref{ex1} can be used to design SM training matrix which attains the MSE lower bound.

Next, we evaluate the channel estimation MSE performances
under different values of multi-paths at Eb/No of $16$
dB. We employ the optimal $(48,22)$-CZCP constructed through Theorem \ref{th6} to generate our SM training matrix. We compare
its channel estimation performance with SM training matrices
from $(16,8)$- CZCP given in \cite{zilong_ccp}, the length-16 GCP given below
\begin{equation}
\left( \begin{matrix}
\mathbf{a} \\
\mathbf{b}
\end{matrix} \right)= \left ( \begin{matrix}
+++++--+++--+-+-\\
+-+-++--+--+++++
\end{matrix} \right),
\end{equation}
and also with a randomly generated sequence, which is generated ``on-the-fly".
\begin{figure}
	\centering
	\includegraphics[width=\columnwidth,draft=false]{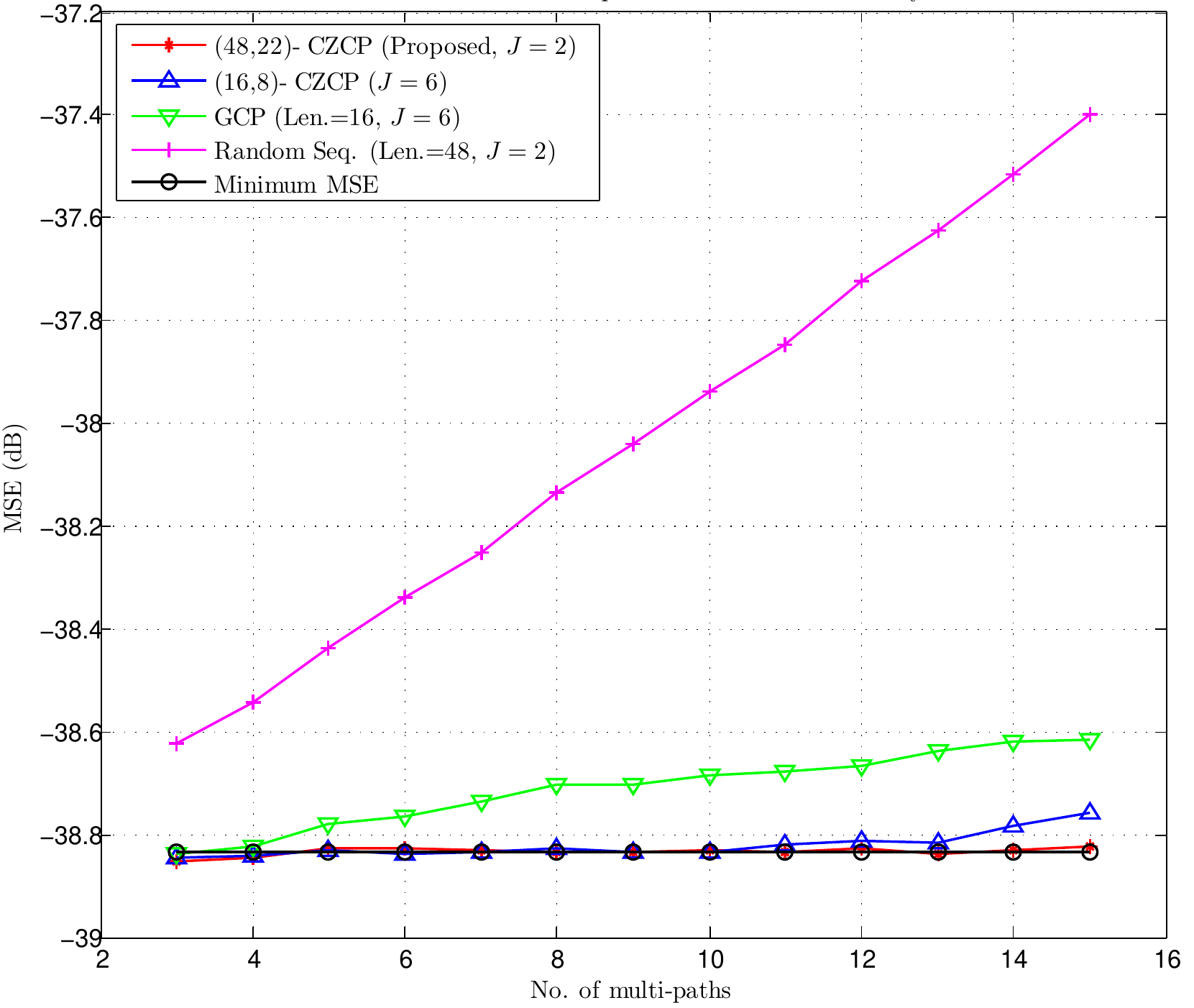}
	\caption{MSE comparison between various sequences, Eb/No=16 dB. \label{final3}}
\end{figure}
Fig. \ref{final3} shows that when the number of multi-paths is $9$ (or less) $(16,8)$- CZCP of \cite{zilong_ccp} can be used to design SM training matrix which achieves the minimum MSEs. $(48,22)$- CZCP, constructed using Theorem \ref{th6} can be used to design SM training matrix which achieves the minimum MSEs when the number of multi-paths is less than or equal to $15$. As per the discussion in \cite{zilong_ccp} $(48,22)$- CZCP can be used to design SM training matrix which achieves the minimum MSEs when the number of multi-paths is less than or equal to $23$.

Hence, in the given conditions, $(48,22)$- CZCP can be used to design optimal training matrix for SM systems using the framework given in \cite{zilong_ccp}.

\section{Concluding Remarks and Open Problems}
In this paper, we have introduced the concept of $CZC_{ratio}$ and re-categorised the CZCPs based on that. We analysed the non-perfect CZCPs and proposed three constructions of non-perfect CZCPs. The first construction is based on GBFs while the second one is based on applying insertion method on the GCPs, which are constructed via Turyn's method. By using GBFs we have constructed CZCPs of lengths $2^{m-1}+2~(m\geq 4)$. By applying insertion method we have constructed CZCPs of lengths $2^\alpha 10^\beta 26^\gamma +2~ (\alpha\geq 1),~10^\beta+2,~26^\gamma+2$ and $10^\beta 26^\gamma +2$. In the final construction, we proposed construction of binary optimal $(12,5)$- CZCPs and $(24,11)$- CZCPs using Barker sequences. These two optimal CZCPs lead to $(12N,5N)$- CZCPs and $(24N,11N)$- CZCPs, where $N$ is the length of a GCP. All these CZCPs can be used to construct cross Z-complementary sets by the method given in \cite{zilong_ccp}. Also, during this work we found one beautiful property of binary $(N,Z)$- CZCPs, stated in Property \ref{pro1}. Through numerical simulations we show that depending on the number of multi-paths our proposed CZCPs can be used to design optimal training sequences for SM systems, based on the framework proposed by Liu \textit{et al.} in \cite{zilong_ccp}.

While calculating the $CZC_{ratio}$ for non-perfect $(N,Z)$- CZCPs, we always take the maximum value of $Z$, i.e., $Z_{\max}=N/2-1$. However, by taking a closer look at \cite[Table I]{zilong_ccp}, we can see that for binary cases, the ``best possible" CZCPs for lengths $18$ and $22$, one can obtain from computer search, have $Z_{\max}$ values $7$ and $9$, respectively. So, it is highly possible to tighten the upper-bound of $Z_{\max}$ for certain lengths.
Along with the above, systematic constructions of optimal CZCPs with new lengths can be considered.

\section*{Appendix A\\Proof of Theorem \ref{th7}}

To prove the theorem, we need the following lemmas.

\begin{lemma}\label{lem5}
	For a binary sequence pair $(\mathbf{c},\mathbf{d})$ of length $N$, $\rho_{\mathbf{c},\overleftarrow{\mathbf{d}}}(\tau)=\rho_{\mathbf{d},\overleftarrow{\mathbf{c}}}(\tau)$.
\end{lemma}
\begin{IEEEproof}
	By definition of ACCF, we have
	\begin{equation}
	\begin{split}
	\rho_{\mathbf{c},\overleftarrow{\mathbf{d}}}(\tau)&=\sum_{i=0}^{N-1-\tau}c_i\overleftarrow{d}_{i+\tau}\\
	&=\sum_{i=0}^{N-1-\tau}c_id_{N-1-i-\tau}\\
	&=\sum_{t=0}^{N-1-\tau}c_{N-1-t-\tau}d_t\\
	&=\rho_{\mathbf{d},\overleftarrow{\mathbf{c}}}(\tau).
	\end{split}
	\end{equation}
\end{IEEEproof}

\begin{lemma}\cite{zilong_ccp}\label{lem6}
	Let $(\mathbf{c},\mathbf{d})$ be a binary $(N,Z)$- CZCP. Then
	\begin{equation}
	\rho_{\mathbf{c},\overleftarrow{\mathbf{c}}}(\tau)-\rho_{\mathbf{d},\overleftarrow{\mathbf{d}}}(\tau)=0, \text{ for all }|\tau|\geq N-Z.
	\end{equation}
\end{lemma}

\subsection*{Proof of Theorem \ref{th7}}
By the Euclidean division theorem, we have $\tau=k_1M+k_2$ where $0\leq k_1<N$ and $0\leq k_2 <M$. By the definition of AACF, we have
\begin{equation}
\begin{split}
&\rho_{\mathbf{e}}(\tau)=\\&\sum_{m=0}^{N-1-k_1}\left[\left(\frac{c_m+d_{N-1-m}}{2}\right)\left(\frac{c_{m+k_1}+d_{N-1-m-k_1}}{2}\right)\rho_{\mathbf{a}}(k_2)\right.\\&\left.+\left(\frac{c_m-d_{N-1-m}}{2}\right)\left(\frac{c_{m+k_1}-d_{N-1-m-k_1}}{2}\right)\rho_{\mathbf{b}}(k_2)+\right.\\ &\left. \left(\frac{c_m+d_{N-1-m}}{2}\right)\left(\frac{c_{m+k_1}-d_{N-1-m-k_1}}{2}\right)\rho_{\mathbf{a},\mathbf{b}}(k_2)+\right.\\ &\left. \left(\frac{c_m-d_{N-1-m}}{2}\right)\left(\frac{c_{m+k_1}+d_{N-1-m-k_1}}{2}\right)\rho_{\mathbf{b},\mathbf{a}}(k_2)+\right.\\& \left. \left(\frac{c_m+d_{N-1-m}}{2}\right)\left(\frac{c_{m+k_1+1}+d_{N-1-m-k_1-1}}{2}\right)\rho_{\mathbf{a}}(k_3)+\right. \\ &\left. \left(\frac{c_m-d_{N-1-m}}{2}\right)\left(\frac{c_{m+k_1+1}-d_{N-1-m-k_1-1}}{2}\right)\rho_{\mathbf{b}}(k_3)+\right.\\ &\left. \left(\frac{c_m+d_{N-1-m}}{2}\right)\left(\frac{c_{m+k_1+1}-d_{N-1-m-k_1-1}}{2}\right)\rho_{\mathbf{b},\mathbf{a}}(k_3)+\right.\\& \left. \left(\frac{c_m-d_{N-1-m}}{2}\right)\left(\frac{c_{m+k_1+1}+d_{N-1-m-k_1-1}}{2}\right)\rho_{\mathbf{a},\mathbf{b}}(k_3)\right],
\end{split}
\end{equation}
where $M-k_2=k_3$.	
Similarly,
\begin{equation}
\begin{split}
&\rho_{\mathbf{f}}(\tau)=\\&\sum_{m=0}^{N-1-k_1}\left[\left(\frac{d_m-c_{N-1-m}}{2}\right)\left(\frac{d_{m+k_1}-c_{N-1-m-k_1}}{2}\right)\rho_{\mathbf{a}}(k_2)\right.\\&\left.+\left(\frac{d_m+c_{N-1-m}}{2}\right)\left(\frac{d_{m+k_1}+c_{N-1-m-k_1}}{2}\right)\rho_{\mathbf{b}}(k_2)+\right.\\ &\left. \left(\frac{d_m-c_{N-1-m}}{2}\right)\left(\frac{d_{m+k_1}+c_{N-1-m-k_1}}{2}\right)\rho_{\mathbf{a},\mathbf{b}}(k_2)+\right.\\ &\left. \left(\frac{d_m+c_{N-1-m}}{2}\right)\left(\frac{d_{m+k_1}-c_{N-1-m-k_1}}{2}\right)\rho_{\mathbf{b},\mathbf{a}}(k_2)+\right.\\& \left. \left(\frac{d_m-c_{N-1-m}}{2}\right)\left(\frac{d_{m+k_1+1}-c_{N-1-m-k_1-1}}{2}\right)\rho_{\mathbf{a}}(k_3)+\right. \\ &\left. \left(\frac{d_m+c_{N-1-m}}{2}\right)\left(\frac{d_{m+k_1+1}+c_{N-1-m-k_1-1}}{2}\right)\rho_{\mathbf{b}}(k_3)+\right.\\ &\left. \left(\frac{d_m-c_{N-1-m}}{2}\right)\left(\frac{d_{m+k_1+1}+c_{N-1-m-k_1-1}}{2}\right)\rho_{\mathbf{b},\mathbf{a}}(k_3)+\right.\\& \left. \left(\frac{d_m+c_{N-1-m}}{2}\right)\left(\frac{d_{m+k_1+1}-c_{N-1-m-k_1-1}}{2}\right)\rho_{\mathbf{a},\mathbf{b}}(k_3)\right],
\end{split}
\end{equation}
where $M-k_2=k_3$.	
Therefore, by some elementary operations and Lemma \ref{lem5}, we have
\begin{equation}\label{eq88}
\begin{split}
&\rho_{\mathbf{e}}(\tau)+\rho_{\mathbf{f}}(\tau)=\\&\frac{1}{4}\left(\rho_{\mathbf{c}}(k_1)+\rho_{\mathbf{d}}(k_1)+\rho_{\overleftarrow{\mathbf{c}}}(k_1)+\rho_{\overleftarrow{\mathbf{d}}}(k_1)\right)\left(\rho_{\mathbf{a}}(k_2)+\rho_{\mathbf{b}}(k_2)\right)\\&+\frac{1}{4}\left(\rho_{\mathbf{c}}(k_1+1)+\rho_{\mathbf{d}}(k_1+1)+\rho_{\overleftarrow{\mathbf{c}}}(k_1+1)+\rho_{\overleftarrow{\mathbf{d}}}(k_1+1)\right)\\&\hspace{4.5cm}\left(\rho_{\mathbf{a}}(M-k_2)+\rho_{\mathbf{b}}(M-k_2)\right)
\end{split}
\end{equation}
When $\tau=MZ+M$, then $k_1=(Z+1)$ and $k_2=0$, therefore from (\ref{eq88}), it is clear that
\begin{equation}
\begin{split}
&\rho_{\mathbf{e}}(MZ+M)+\rho_{\mathbf{f}}(MZ+M)=\\&\frac{1}{2}\left(\rho_{\mathbf{c}}(Z+1)+\rho_{\mathbf{d}}(Z+1)\right)\left(\rho_{\mathbf{a}}(0)+\rho_{\mathbf{b}}(0)\right).
\end{split}
\end{equation}
otherwise, for all other values of $\tau$,
\begin{equation}
\rho_{\mathbf{e}}(\tau)+\rho_{\mathbf{f}}(\tau)=0.
\end{equation}

Now to calculate the cross-correlation, for $M-k_2=k_3$, we have
\begin{equation}\label{eq97}
\begin{split}
&\rho_{\mathbf{e},\mathbf{f}}(\tau)=\\&\sum_{m=0}^{N-1-k_1}\left[\left(\frac{c_m+d_{N-1-m}}{2}\right)\left(\frac{d_{m+k_1}-c_{N-1-m-k_1}}{2}\right)\rho_{\mathbf{a}}(k_2)\right.\\&\left.+\left(\frac{c_m-d_{N-1-m}}{2}\right)\left(\frac{d_{m+k_1}+c_{N-1-m-k_1}}{2}\right)\rho_{\mathbf{b}}(k_2)+\right.\\ &\left. \left(\frac{c_m+d_{N-1-m}}{2}\right)\left(\frac{d_{m+k_1}+c_{N-1-m-k_1}}{2}\right)\rho_{\mathbf{a},\mathbf{b}}(k_2)+\right.\\ &\left. \left(\frac{c_m-d_{N-1-m}}{2}\right)\left(\frac{d_{m+k_1}-c_{N-1-m-k_1}}{2}\right)\rho_{\mathbf{b},\mathbf{a}}(k_2)+\right.\\& \left. \left(\frac{c_m+d_{N-1-m}}{2}\right)\left(\frac{d_{m+k_1+1}-c_{N-1-m-k_1-1}}{2}\right)\rho_{\mathbf{a}}(k_3)+\right. \\ &\left. \left(\frac{c_m-d_{N-1-m}}{2}\right)\left(\frac{d_{m+k_1+1}+c_{N-1-m-k_1-1}}{2}\right)\rho_{\mathbf{b}}(k_3)+\right.\\ &\left. \left(\frac{c_m+d_{N-1-m}}{2}\right)\left(\frac{d_{m+k_1+1}+c_{N-1-m-k_1-1}}{2}\right)\rho_{\mathbf{b},\mathbf{a}}(k_3)+\right.\\& \left. \left(\frac{c_m-d_{N-1-m}}{2}\right)\left(\frac{d_{m+k_1+1}-c_{N-1-m-k_1-1}}{2}\right)\rho_{\mathbf{a},\mathbf{b}}(k_3)\right].
\end{split}
\end{equation}	
Similarly,
\begin{equation}\label{eq98}
\begin{split}
&\rho_{\mathbf{f},\mathbf{e}}(\tau)=\\&\sum_{m=0}^{N-1-k_1}\left[\left(\frac{d_m-c_{N-1-m}}{2}\right)\left(\frac{c_{m+k_1}+d_{N-1-m-k_1}}{2}\right)\rho_{\mathbf{a}}(k_2)\right.\\&\left.+\left(\frac{d_m+c_{N-1-m}}{2}\right)\left(\frac{c_{m+k_1}-d_{N-1-m-k_1}}{2}\right)\rho_{\mathbf{b}}(k_2)+\right.\\ &\left. \left(\frac{d_m-c_{N-1-m}}{2}\right)\left(\frac{c_{m+k_1}-d_{N-1-m-k_1}}{2}\right)\rho_{\mathbf{a},\mathbf{b}}(k_2)+\right.\\ &\left. \left(\frac{d_m+c_{N-1-m}}{2}\right)\left(\frac{c_{m+k_1}+d_{N-1-m-k_1}}{2}\right)\rho_{\mathbf{b},\mathbf{a}}(k_2)+\right.\\& \left. \left(\frac{d_m-c_{N-1-m}}{2}\right)\left(\frac{c_{m+k_1+1}+d_{N-1-m-k_1-1}}{2}\right)\rho_{\mathbf{a}}(k_3)+\right. \\ &\left. \left(\frac{d_m+c_{N-1-m}}{2}\right)\left(\frac{c_{m+k_1+1}-d_{N-1-m-k_1-1}}{2}\right)\rho_{\mathbf{b}}(k_3)+\right.\\ &\left. \left(\frac{d_m-c_{N-1-m}}{2}\right)\left(\frac{c_{m+k_1+1}-d_{N-1-m-k_1-1}}{2}\right)\rho_{\mathbf{b},\mathbf{a}}(k_3)+\right.\\& \left. \left(\frac{d_m+c_{N-1-m}}{2}\right)\left(\frac{c_{m+k_1+1}+d_{N-1-m-k_1-1}}{2}\right)\rho_{\mathbf{a},\mathbf{b}}(k_3)\right].
\end{split}
\end{equation}	

Simplifying (\ref{eq97}), (\ref{eq98}) and using Lemma \ref{lem6}, we get for $ZM\leq \tau <NM$,
\begin{equation}
\rho_{\mathbf{e},\mathbf{f}}(\tau)+\rho_{\mathbf{f},\mathbf{e}}(\tau)=0,
\end{equation}
since, for $ZM\leq \tau <NM$, we have $Z\leq k_1<N$. Hence, $(\mathbf{e},\mathbf{f})$ is a $(NM,ZM)$- CZCP. This completes the proof.


\begin{thebibliography}{20}
	\bibitem{golay1}
	M. J. E. Golay, ``Static multislit spectrometry and its application to the
	panoramic display of infrared spectra," \emph{J. Opt. Soc. Am.}, vol. 41, no. 7,
	pp. 468-472, Jul. 1951.
	
	\bibitem{golay2}
	M. J. E. Golay, ``Complementary series," \emph{IRE Trans. Inf. Theory}, vol. 7, no. 2, pp.
	82-87, Apr. 1961.
	
	\bibitem{Borwein03}
	P. B. Borwein and R. A. Ferguson, ``A complete description of Golay
	pairs for lengths up to 100," \emph{Math. Comput.}, vol. 73, no. 246, pp. 967-985, Jul. 2003.
	
	\bibitem{tseng72}
	C.-C. Tseng and C. L. Liu, ``Complementary sets of sequences," \emph{IEEE
	Trans. Inf. Theory}, vol. 18, no. 5, pp. 644-652, Sep. 1972.

\bibitem{spasojevic}
P. Spasojevic and C. N. Georghiades, ``Complementary sequences for
ISI channel estimation," \emph{IEEE Trans. Inf. Theory}, vol. 47, no. 3, pp.
1145-1152, Mar. 2001.

	
\bibitem{popovic}
B. M. Popovic, ``Synthesis of power efficient multitone signals with
flat amplitude spectrum," \emph{IEEE Trans. Commun.}, vol. 39, no. 7,
pp. 1031-1033, Jul. 1991.

\bibitem{davis99}
J. A. Davis and J. Jedwab, ``Peak-to-mean power control in OFDM,
Golay complementary sequences, and Reed-Muller codes," \emph{IEEE Trans.
	Inf. Theory}, vol. 45, no. 7, pp. 2397-2417, Nov. 1999.

\bibitem{paterson}
K. G. Paterson, ``Generalized Reed-Muller codes and power control
in OFDM modulation," \emph{IEEE Trans. Inf. Theory}, vol. 46, no. 1,
pp. 104-120, Jan. 2000.


\bibitem{schmidt}
K.-U. Schmidt, ``Complementary sets, generalized Reed-Muller codes,
and power control for OFDM," \emph{IEEE Trans. Inf. Theory}, vol. 53, no. 2,
pp. 808-814, Feb. 2007.	

\bibitem{fan}
P. Fan, W. Yuan, Y. Tu, ``Z-complementary binary sequences," \emph{IEEE Signal Process. Lett.} vol. 14 no. 8 pp. 509-512 Aug. 2007.


\bibitem{li}
X. Li, P. Fan, X. Tang, Y. Tu, ``Existence of binary Z-complementary pairs," \emph{IEEE Signal Process. Lett.} vol. 18 no. 1 pp. 63-66 Jan. 2011.

\bibitem{zilongobzcp}
Z. Liu, U. Parampalli, Y. L. Guan, ``Optimal odd-length binary Z-complementary pairs," \emph{IEEE Trans. Inf. Theory} vol. 60 no. 9 pp. 5768-5781 Sep. 2014.

\bibitem{zilongebzcp}
Z. Liu, U. Parampalli, Y. L. Guan, ``On even-period binary Z-complementary pairs with large ZCZs," \emph{IEEE Signal Process. Lett.} vol. 21 no. 3 pp. 284-287 Mar. 2014.

\bibitem{chen17}
C.-Y. Chen, ``A novel construction of Z-complementary pairs based on generalized Boolean functions," \emph{IEEE Signal Process. Lett.} vol. 24 no. 7 pp. 987-990 Jul. 2017.

\bibitem{Avik_iwsda}
A. R. Adhikary, S. Majhi, Z. Liu, and Y. L. Guan, ``New optimal binary Z-complementary pairs of odd lengths," \emph{in Proc. The 8th IEEE Int. Workshop on Signal Design and its App. in Commun. (IWSDA)}, Sapporo, Japan, Sept. 2017.

\bibitem{Avik_tit}
A. R. Adhikary, S. Majhi, Z. Liu and Y. L. Guan, ``New sets of optimal odd-length binary Z-complementary pairs," \emph{IEEE Trans. Inf. Theory}, vol. 66, no. 1, pp. 669-678, Jan. 2020.

\bibitem{Avik_ebzcp}
A. R. Adhikary, S. Majhi, Z. Liu and Y. L. Guan, ``New sets of even-length binary Z-complementary pairs with asymptotic ZCZ ratio of $3/4$," \emph{IEEE Signal Process. Lett.}, vol. 25, no. 7, pp. 970-973, Jul. 2018.

\bibitem{Mesleh1}
R. Mesleh, H. Haas, Y. Lee, and S. Yun, ``Interchannel interference
avoidance in MIMO transmission by exploiting spatial information," in
\emph{Proc. 16th IEEE Int. Symp. PIMRC}, Berlin, Germany, 2005, vol. 1, pp. 141-145.

\bibitem{Mesleh2}
R. Mesleh, H. Haas, S. Sinanovi, C. W. Ahn, and S. Yun, ``Spatial
modulation," \emph{IEEE Trans. Veh. Technol.}, vol. 57, no. 4, pp. 2228-2241, Jul. 2008.

\bibitem{renzo}
M. Di Renzo, H. Haas, and P. M. Grant, ``Spatial modulation for multiple antenna wireless systems: A survey," \emph{IEEE Commun. Mag.}, vol. 49, no.12, pp. 182-191, Dec. 2011.

\bibitem{yang}
P. Yang, M. D. Renzo, Y. Xiao, S. Li, and L. Hanzo, ``Design guidlines
for spatial modulation," \emph{IEEE Commun. Surveys Tuts.}, vol. 17, no. 1, pp. 6-26, First Quart., 2015.

\bibitem{wen}
M. Wen, B. Zheng, K. Kim, M. Renzo, T. Tsiftsis, K.-C. Chen, and N.
Al-Dhahir, ``A survey on spatial modulation in emerging wireless systems: research progresses and applications," \emph{IEEE J. Sel. Areas Commun.}, vol. 37, no. 9, pp. 1949-1972, Sep. 2019.


\bibitem{yang1}
P. Yang, Y. Xiao, Y. L. Guan, K. V. S. Hari, A. Chockalingam, S.
Sugiura, H. Hass, M. D. Renzo, C. Masouros, Z. Liu, L. Xiao, S. Li, and
L. Hanzo, ``Single-carrier SM-MIMO: a promising design for broadband
large-scale antenna systems," \emph{IEEE Commun. Surveys Tuts.}, vol. 18, no. 3, pp. 1687-1716, Third Quart., 2016.

\bibitem{Jeganathan}
J. Jeganathan, A. Ghrayeb, and L. Szczecinski, ``Spatial modulation:
Optimal detection and performance analysis," \emph{IEEE Commun. Lett.}, vol. 12, no. 8, pp. 545-547, Aug. 2008.

\bibitem{Renzo12}
M. D. Renzo and H. Haas, ``Bit error probability of SM-MIMO over
generalized fading channels," \emph{IEEE Trans. Veh. Technol.}, vol. 61, no. 3, pp. 1124-1144, Mar. 2012.




\bibitem{yang3}
S. A. Yang and J. Wu, ``Optimal binary training sequence design for
multiple-antenna systems over dispersive fading channels," \emph{IEEE Trans. Veh. Technol.}, vol. 51, pp. 1271-1276, Sept. 2002.

\bibitem{Fragouli}
C. Fragouli, N. Al-Dhahir, and W. Turin, ``Training-based channel
estimation for multiple-antenna broadband transmissions," \emph{IEEE Trans. Wireless Commun.}, vol. 2, no. 2, pp. 384-391, Mar. 2003.


\bibitem{fan2}
P. Fan and W. H. Mow, ``On optimal training sequence design for
multiple-antenna systems over dispersive fading channels and its extensions," \emph{IEEE Trans. Veh. Technol.}, vol. 53, no. 5, pp. 1623-1626, Sep. 2004.

\bibitem{Sugiura}
S. Sugiura and L. Hanzo, ``Effects of channel estimation on spatial
modulation," \emph{IEEE Signal Process. Lett.}, vol. 19, no. 12, pp. 805-808, Dec. 2012.






\bibitem{zilong_ccp}
Z. Liu, P. Yang, Y. L. Guan, P. Xiao, ``Cross Z-complementary pairs for optimal training in spatial modulation over frequency selective channels," \emph{IEEE Trans. Signal Process.}, vol. 68, pp. 1529-1543, 2020.

\bibitem{rathinakumar}
A. Rathinakumar and A. K. Chaturvedi, ``Complete mutually orthogonal Golay complementary sets from Reed-Muller codes" \emph{IEEE Trans. Inf. Theory} vol. 54 no. 3 pp. 1339-1346 Mar. 2008.

\bibitem{Turyn74}
R. Turyn, ``Hadamard matrices, Baumert-Hall units, four-symbol sequences,
pulse compression and surface wave encodings," \emph{J. Combin.
Theory (A)}, vol. 16, pp. 313-333, 1974.



\bibitem{fan_book}
P. Fan and M. Darnell, Sequence Design for Communications Applications. New York, NY, USA: Wiley, 1996.

\bibitem{chen2016}
C. Chen, ``Complementary sets of non-power-of-two length for peak-to-average power ratio reduction in OFDM," \emph{IEEE Trans. Inf. Theory}, vol. 62, no. 12, pp. 7538-7545, Dec. 2016.


\end{thebibliography}
	\end{document}